\documentclass[12pt, a4paper]{article}
\pdfoutput=1
\usepackage[utf8]{inputenc}
\usepackage{fancybox}
\usepackage[T1]{fontenc}
\usepackage{lmodern, textcomp}
\usepackage{bm}
\usepackage[american]{babel}
\usepackage{csquotes}

\usepackage{eurosym}
\usepackage[nottoc]{tocbibind}
\usepackage{authblk}
\usepackage{fancyhdr}
\usepackage{xspace}
\usepackage{soul}

\usepackage{graphicx}
\usepackage{subcaption}
\usepackage{float}
\usepackage{makecell}
\usepackage{multirow}
\usepackage{multicol} 
\usepackage{marginnote}
\usepackage[multiple]{footmisc}

\usepackage[usenames, dvipsnames, svgnames, table]{xcolor}

\usepackage[backend=bibtex8,
            style=ieee,
            dashed=false,
            citestyle=numeric-comp,
            maxbibnames=99,
			sorting=none,
            sortcase=false,
            sortcites=true,
            giveninits=true]{biblatex}

\usepackage{amsmath, amsfonts, amssymb}
\usepackage{mathtools}
\usepackage{mathrsfs}
\usepackage{cancel}
\usepackage{braket}
\usepackage{tensor}
\usepackage{slashed}
\usepackage{siunitx}
\usepackage{listings}

\usepackage{stmaryrd}
\usepackage{caption}
\usepackage{relsize,exscale}
\usepackage{array}
\usepackage[toc,page]{appendix}

\usepackage{enumerate}

\DeclareSymbolFont{largesymbols}{OMX}{cmex}{m}{n}

\setcellgapes{1pt}
\makegapedcells
\newcolumntype{R}[1]{>{\raggedleft\arraybackslash }b{#1}}
\newcolumntype{L}[1]{>{\raggedright\arraybackslash }b{#1}}
\newcolumntype{C}[1]{>{\centering\arraybackslash }b{#1}}

\newcommand{\Tr}{\mathrm{Tr}}

\newtheorem{remark}{Remark}

\newcommand{\beq}{\begin{equation}}
\newcommand{\eeq}{\end{equation}}
\newcommand{\bea}{\begin{eqnarray}}
\newcommand{\eea}{\end{eqnarray}}
\definecolor{mygray}{gray}{0.3}

\newcommand{\bes}{\begin{eqnarray}}
\newcommand{\ees}{\end{eqnarray}}

\newcommand\restr[2]{{
  \left.\kern-\nulldelimiterspace 
  #1 
  \vphantom{\big|} 
  \right|_{#2} 
  }}
\newcommand{\extd}[1]{\mathrm {d}{#1}}

\newcommand{\intpk}{\left(\int_0^k \extd{p}\, p\, \rho(p^2)\right)}
\usepackage{multicol}
\usepackage{amssymb}

\newcommand{\eqdef}{\overset{\text{def}}{=\joinrel=}}

\usepackage[heightrounded, top=3.5cm, bottom=3.5cm, left=3cm, right=3cm, headheight=14pt]{geometry}

\usepackage[pdftex, breaklinks=true, linktocpage,
	colorlinks=true, urlcolor=blue, linkcolor=blue, citecolor=red
]{hyperref}

\newcommand{\email}[1]{\href{mailto:#1}{\nolinkurl{#1}}}
\newcommand{\emailfoot}[1]{\thanks{\email{#1}}}

\usepackage{marginnote}
\newcounter{draftcommentcnt}
\NewDocumentCommand{\draftcomment}{s O{red} m}{%
	\def\margnote{\IfBooleanTF{#1}{\marginnote}{\marginpar}}%
	\stepcounter{draftcommentcnt}%
	\textcolor{#2}{#3}%
	\margnote{\textcolor{#2}{$\Leftarrow$ \arabic{draftcommentcnt}}}%
}

\fancypagestyle{preprint}{
	\fancyhf{}

	\fancyhead[R]{\textsf{\preprint}}
}

\numberwithin{equation}{section}

\hypersetup{
	pdftitle={Functional renormalization group for signal detection and stochastic ergodicity breaking},
}
\title{Functional renormalization group for signal detection and stochastic ergodicity breaking}
\author[1,2,3]{Harold Erbin\emailfoot{erbin@mit.edu}}
\author[1]{Riccardo Finotello\emailfoot{riccardo.finotello@cea.fr}}
\author[1,4]{Bio Wahabou Kpera\emailfoot{wahaboukpera@gmail.com}}
\author[1]{Vincent Lahoche\emailfoot{vincent.lahoche@cea.fr}}
\author[1,4]{Dine Ousmane Samary\emailfoot{dine.ousmanesamary@cipma.uac.bj}}

\affil[1]{%
	Université Paris-Saclay, \textsc{Cea}, \textsc{List}, Palaiseau, F-91120, France
}

\affil[2]{%
	Center for Theoretical Physics, Massachusetts Institute of Technology
	\protect\\
	Cambridge, MA 02139, USA
}

\affil[3]{%
	NSF AI Institute for Artificial Intelligence and Fundamental Interactions
}

\affil[4]{%
	Faculté des Sciences et Techniques (ICMPA-UNESCO Chair)
	\protect\\
	Université d'Abomey-Calavi, 072 BP 50, Benin
}

\begin{document}

\maketitle

\begin{abstract}
Signal detection is one of the main challenges of data science.
As it often happens in data analysis, the signal in the data may be corrupted by noise.
There is a wide range of techniques aimed at extracting the relevant degrees of freedom from data.
However, some problems remain difficult.
It is notably the case of signal detection in almost continuous spectra when the signal-to-noise ratio is small enough.
This paper follows a recent bibliographic line which tackles this issue with field-theoretical methods.
Previous analysis focused on equilibrium Boltzmann distributions for some effective field representing the degrees of freedom of data.
It was possible to establish a relation between signal detection and $\mathbb{Z}_2$-symmetry breaking.
In this paper, we consider a stochastic field framework inspiring by the so-called ``Model A'', and show that the ability to reach or not an equilibrium state is correlated with the shape of the dataset.
In particular, studying the renormalization group of the model, we show that the weak ergodicity prescription is always broken for signals small enough, when the data distribution is close to the Marchenko-Pastur (MP) law.
This, in particular, enables the definition of a detection threshold in the regime where the signal-to-noise ratio is small enough.
\end{abstract}

\noindent \textbf{keywords:} Functional renormalization group, Stochastic field theory, Signal detection, random matrix theory.

\setcounter{footnote}{0}
\newpage

\hrule
\pdfbookmark[1]{\contentsname}{toc}
\tableofcontents
\bigskip
\hrule

\clearpage


\section{Introduction}

In the recent years, many authors pointed out the connection between the Renormalization Group (RG) and data science -- see for instance~\cite{Li_2018,De_Mello_Koch_2020,Koch_Janusz_2018,mehta2014exact,Bradde_2017,koch2020unsupervised,Koch_2020,Halverson_2021,grosvenor2021edge,erbin2021nonperturbative,maiti2021symmetry,erbin2022renormalization,banta2023structures,grosvenor2022edge,kline2023multi} and references therein.
This is not surprising, as for both techniques the goal is the same: Extract large-scale regularities and relevant features for a system with interacting (i.e.\ highly correlated) degrees of freedom.
Indeed, the RG aims to track a small number of relevant parameters (``couplings''), which describe the effective long-distance physics, in a quantum or statistical system involving a large number of interacting degrees of freedom (like a field theory).
There are many incarnations of this idea, the most popular nowadays being the Wilsonian point of view~\cite{wilson1983renormalization}.
In this realization, RG transformations look as partial integration of fluctuations at the microscopic scale, that leave the long-distance physics unchanged but modify the effective couplings between infrared degrees of freedom.
The basic incarnation of this strategy is Kadanoff's block spin construction, where, at each step, the effective spin variables are locally replaced by their average.
Thanks to its rather general scope, RG finds applications in almost all physical domains, see~\cite{Dupuis_2021} and reference therein for a general overview of RG applications, and~\cite{ZinnJustinBook1,ZinnJustinBook2} for a comprehensive presentation of concepts.

\begin{figure}[t]
    \centering
    \begin{subfigure}[t]{0.49\linewidth}
        \centering
        \includegraphics[width=\linewidth]{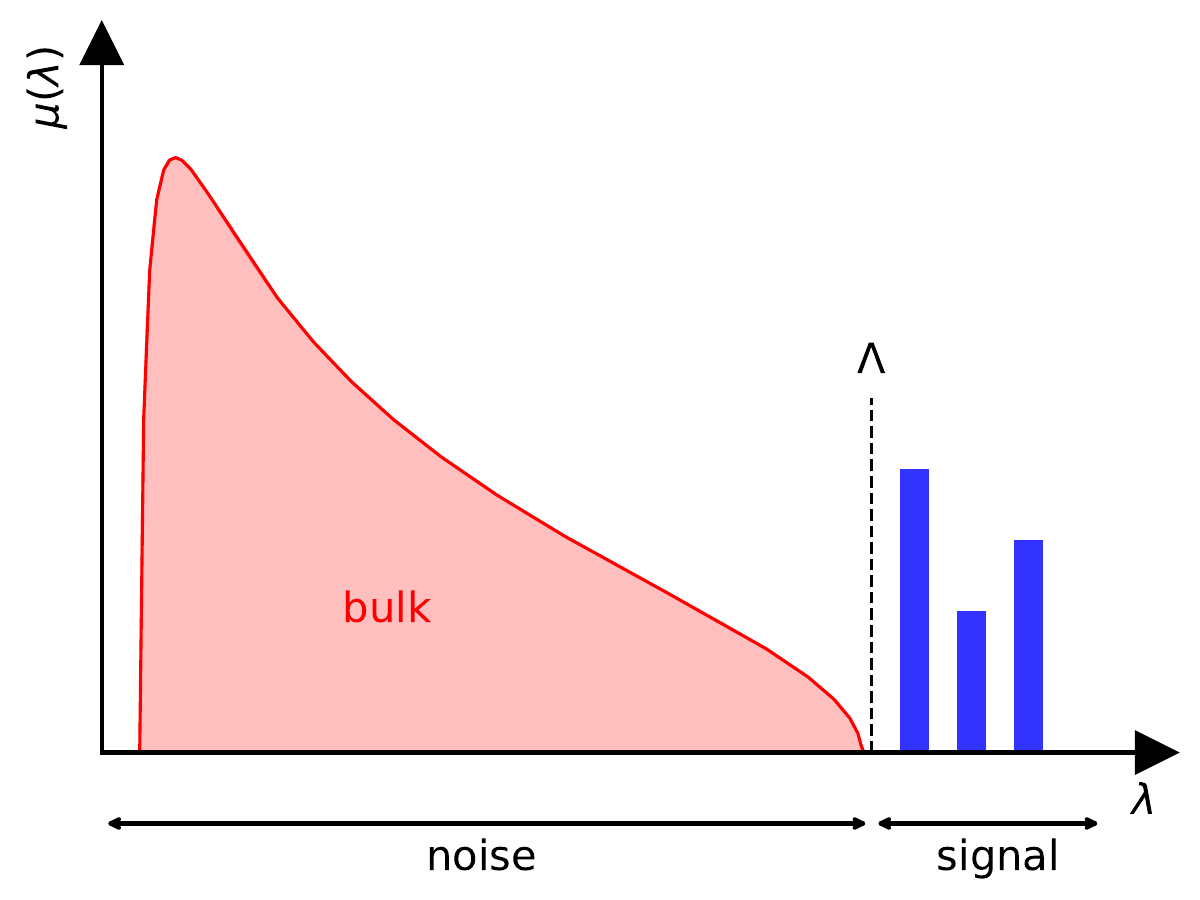}
        \caption{%
            Spectrum with localized spikes.
        }
    \end{subfigure}
    \hfill
    \begin{subfigure}[t]{0.49\linewidth}
        \centering
        \includegraphics[width=\linewidth]{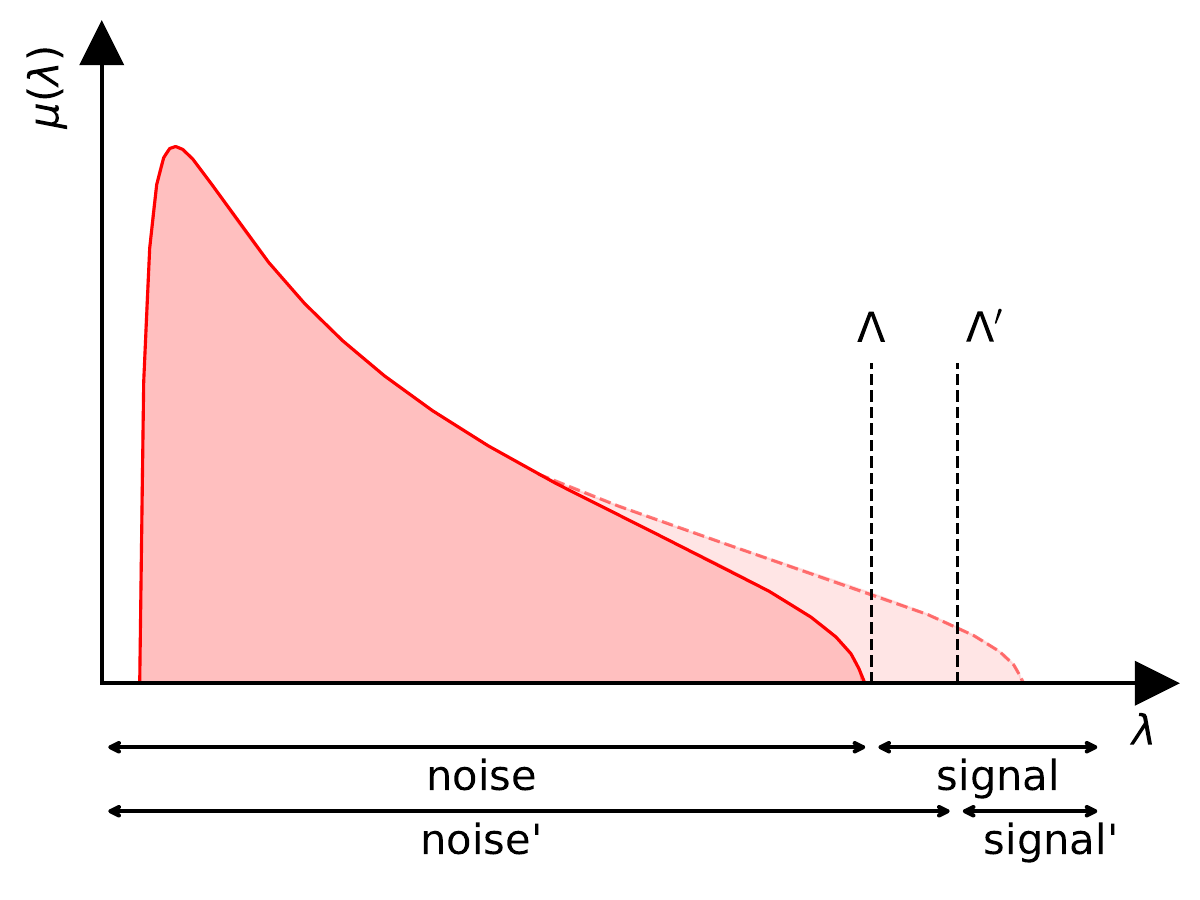}
        \caption{%
            Spectrum with continuous deformation.
        }
    \end{subfigure}
    \caption{%
        Depending on the nature of the underlying data, an empirical spectrum can exhibit some localized spikes (left) out of a bulk (i.e.\ noise, in red) made of delocalized eigenvectors (i.e.\ relevant information, in blue), in which case the cut-off $\Lambda$ provides a clean separation between delocalized eigenvectors and localized ones.
        In the case of nearly continuous spectra (right), the position of the cut-off $\Lambda$ is arbitrary, and the separation of the signal can become impossible.
    }
    \label{fig1}
\end{figure}

In some cases, the links between data science methods and RG may look like a formal analogue rather than a guiding principle.
Recently, the point of view seems to be evolving, and several authors have started to take very seriously the idea that there exists a deeper connection between the two.
Moreover, the idea that some problems inherent to big data analysis, and to artificial intelligence (AI) in general, can be approached as physical systems is starting to gain ground.
One can mention for instance the series of papers~\cite{erbin2022renormalization,maiti2021symmetry,erbin2021nonperturbative,Halverson_2021,grosvenor2021edge} aiming at building an effective field theory model for artificial neural networks, and at studying their properties analysing the RG.
Other approaches focus on a comparison with so-called ``explainable'' methods~\cite{general1}, such as Principal Components Analysis (PCA)~\cite{B_ny_2018,B_ny_2015,Bradde_2017,Bial1,Bial2}.
This paper follows the bibliographic line of the authors~\cite{lahoche2020generalized,lahoche2020field,lahoche2021signal,lahoche2020field2,LahocheSignal2022,} and is part of this dynamic.
These works focus in particular on a problem where the current analysis methods often fail: signal detection in nearly continuous spectra\footnote{When only a few isolated spikes exist, detection is easy and general theorems exist, viewing the detection as a phase transition -- the so-called ``BBP'' theorem, see~\cite{BBP}.} (see Figure~\ref{fig1}).

\begin{figure}[t]
    \centering
    \begin{subfigure}[t]{0.32\linewidth}
        \centering
        \includegraphics[width=\linewidth]{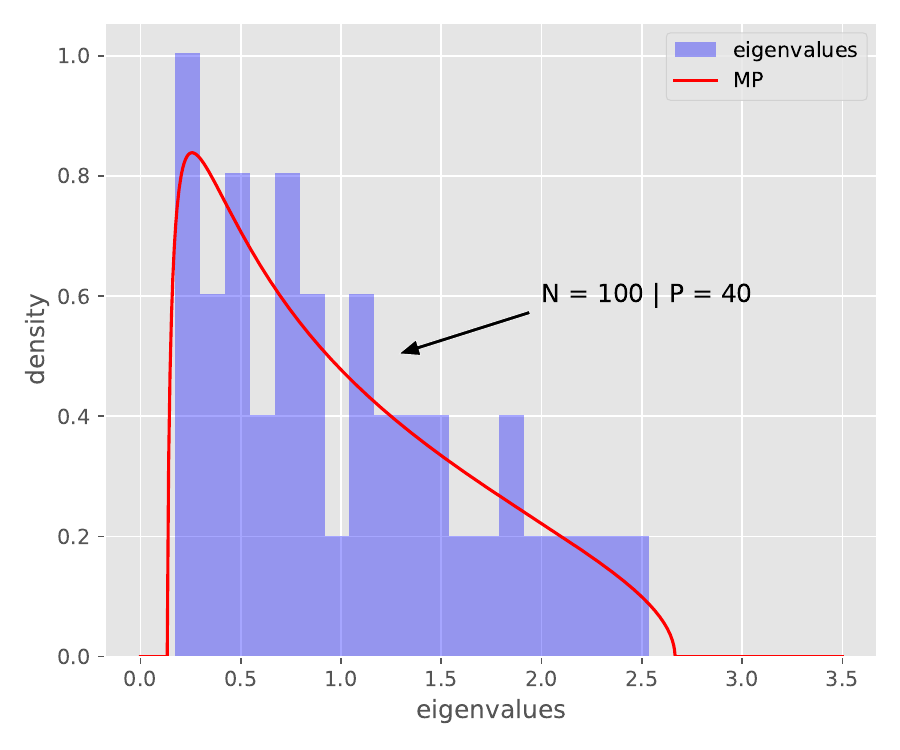}
        \caption{%
            $N = 100$, $P = 40$.
        }
    \end{subfigure}
    \hfill
    \begin{subfigure}[t]{0.32\linewidth}
        \centering
        \includegraphics[width=\linewidth]{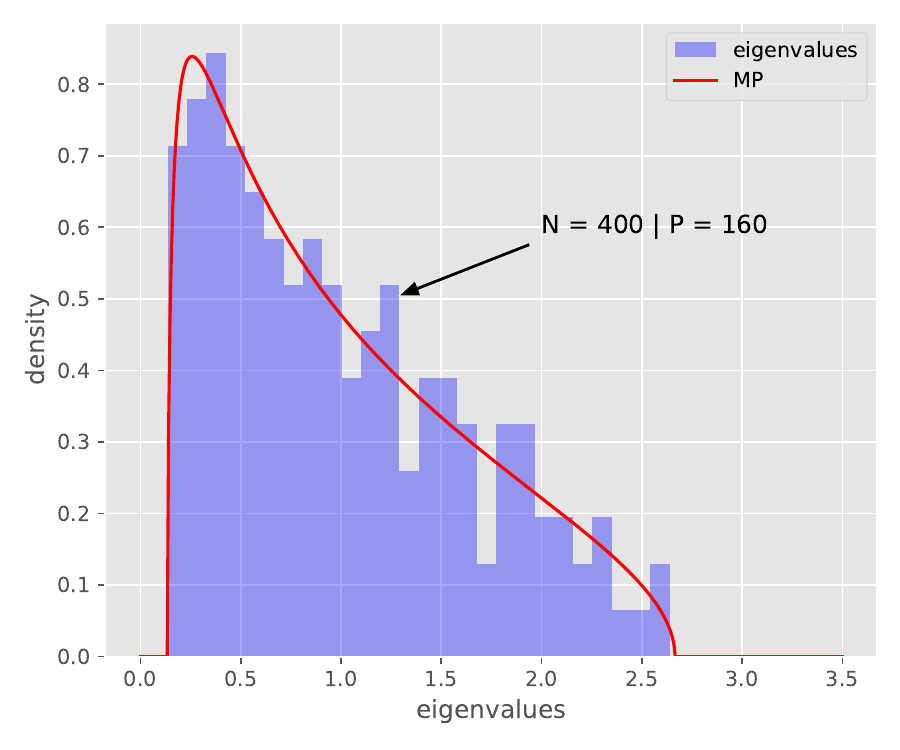}
        \caption{%
            $N = 400$, $P = 160$.
        }
    \end{subfigure}
    \hfill
    \begin{subfigure}[t]{0.32\linewidth}
        \centering
        \includegraphics[width=\linewidth]{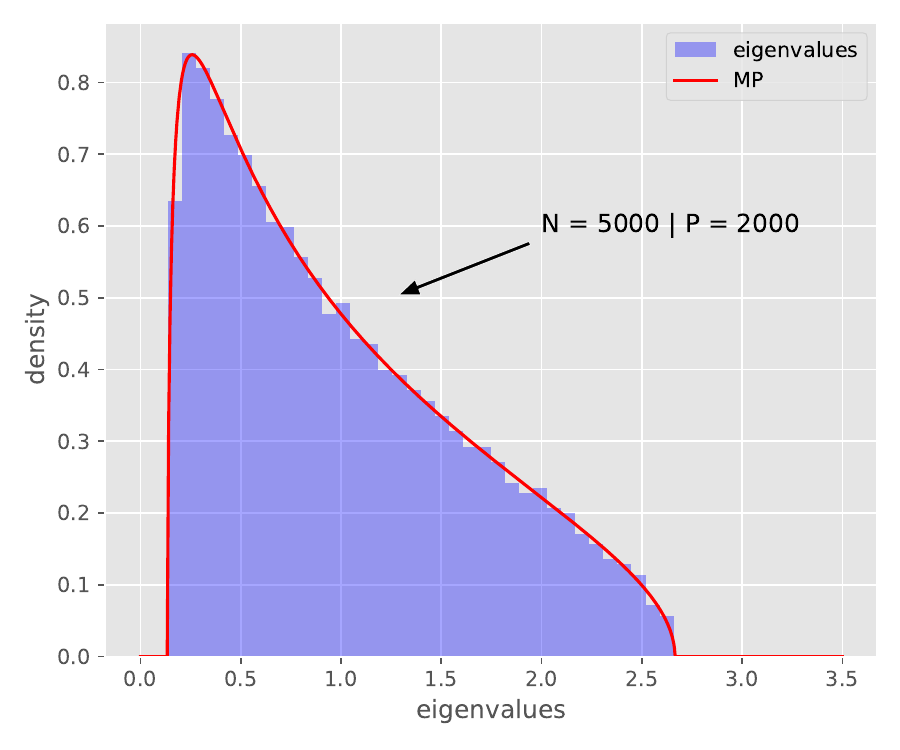}
        \caption{%
            $N = 5000$, $P = 2000$.
        }
    \end{subfigure}
    \caption{%
        Convergence toward Marchenko-Pastur law for large size $N \times P$ Wishart matrices ($P / N = 0.4$).
    }
    \label{fig2}
\end{figure}

The idea behind this approach is based on the concept of universality.
Totally noisy spectra have indeed a universal character, close for instance to properties of large size Wishart matrices, whose spectra follow the Marchenko-Pastur law, see Figure~\ref{fig2}.
In other words, the spectra observed for high-dimensional data are usually completely ``blind'' to the real nature of the degrees of freedom involved in these statistics, whether it is the activity of biological neural networks~\cite{neuroscience}, financial data~\cite{finance1}, correlations between genes in DNA and so on.
Hence, the path proposed in~\cite{lahoche2020generalized,lahoche2020field,lahoche2021signal,lahoche2020field2,LahocheSignal2022,} leads to the idea that the problem of signal detection can be equivalent to the RG study.
We can then design an analogue statistical field theory model as follows.
Suppose we are able, for a particular problem whose large-scale statistics are in the neighbourhood of a matrix universality class, to propose an effective field theory model exploiting the specific nature of the considered degrees of freedom.
Furthermore, suppose that this field theory says something about the presence or absence of the signal.
Hence, this same field theory must be able to give the same kind of answer for any data in the neighbourhood of the same universality class.

\begin{figure}[t]
    \centering
    \begin{subfigure}[t]{0.49\linewidth}
        \centering
        \includegraphics[width=\linewidth]{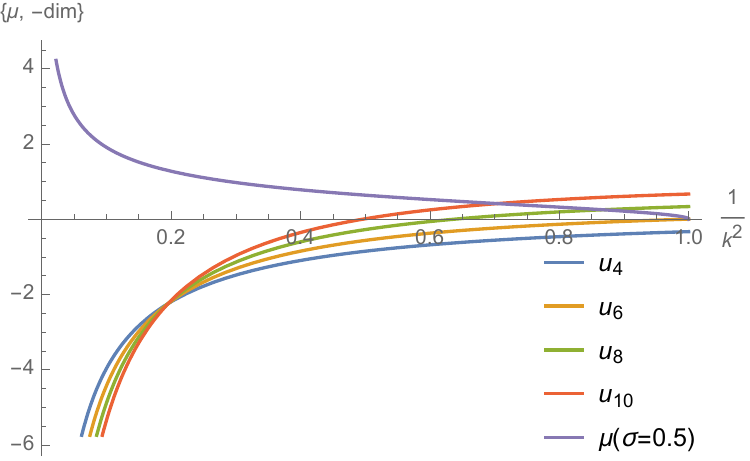}
        \caption{%
            Canonical dimensions.
        }
    \end{subfigure}
    \hfill
    \begin{subfigure}[t]{0.49\linewidth}
        \centering
        \includegraphics[width=\linewidth]{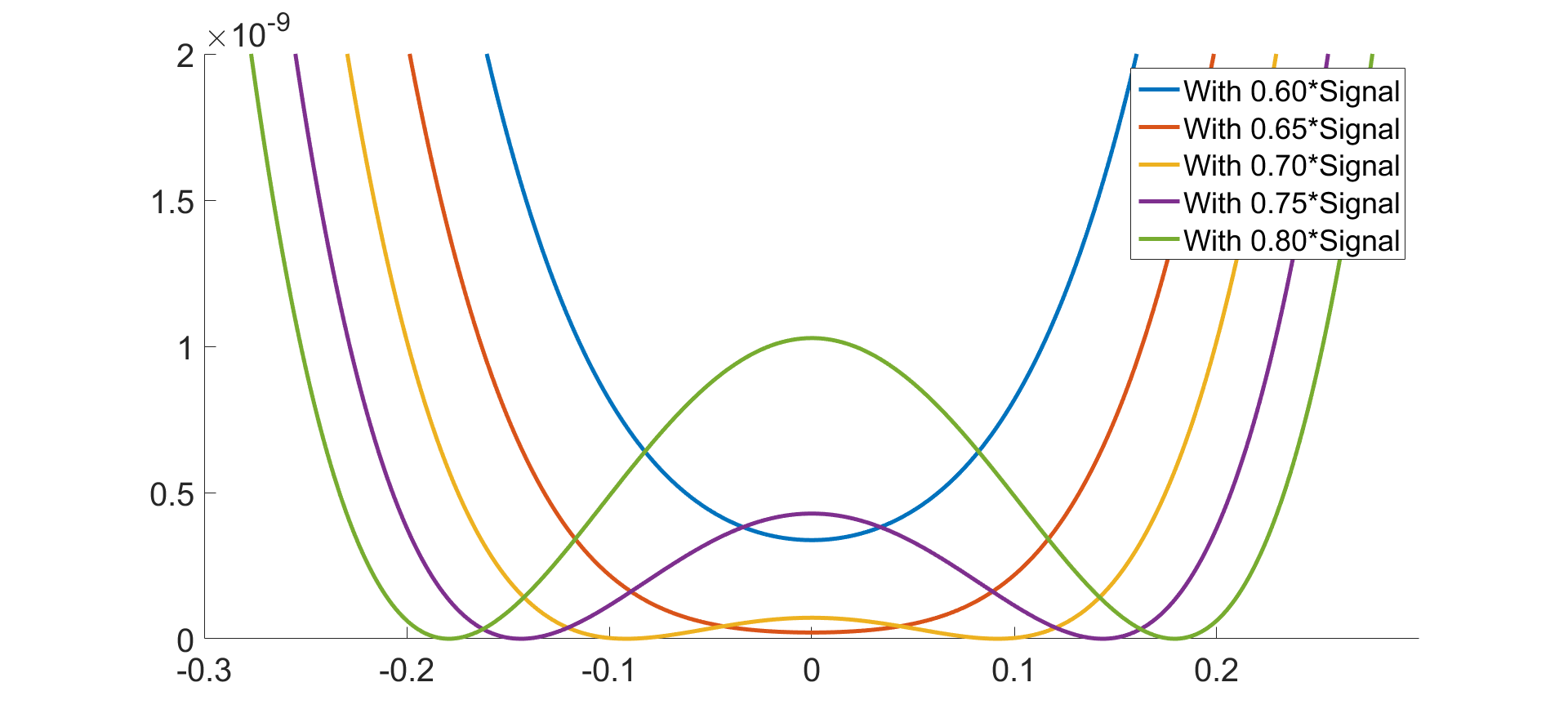}
        \caption{%
            IR effective potential.
        }
    \end{subfigure}
    \caption{%
        Behaviour of the canonical dimensions (left) of local couplings ($u_{2n}$) of $\phi^{2n}$ interaction for Marchenko-Pastur law and shape of the effective potential in the deep IR (right), depending on the strength of the signal embedded in a random Wishart matrix.
        Pictures taken from~\cite{LahocheSignal2022}.
    }
    \label{canonical-dimensions}
\end{figure}

The usual example is an Ising-type model, for which one can easily build an effective field theory once the moments of order $1$ and $2$ are fixed.
The minimal model (in the sense of ``less structured'': pioneer works on this topic of maximum entropy inference in physics are those of Jaynes~\cite{Jaynes1,Jaynes2}) for this field theory is the one of maximum entropy, which takes the form of a Boltzmann law $\rho_{\text{eq}}[\phi]\sim e^{-\mathcal{S}[\phi]}$ for the field $\phi$, with a non-trivial Gaussian measure such that $2$-point function $G$ is identified with the covariance matrix $\mathcal{C}$ of the dataset.
We thus proposed an effective field theory, general in scope, and unconventional in the construction of the RG due to the specific nature of the Gaussian kernel spectrum.
This field theory describes an abstract type of matter filling a one-dimensional space, and whose interacting two-point density spectrum is the data covariance matrix.
The reference~\cite{LahocheSignal2022} is a general and comprehensive review of the state of the art of this approach.
In summary, we can make the following empirical statements about RG analysis in the vicinity of the Marchenko-Pastur law (see Figure~\ref{canonical-dimensions}):

\begin{itemize}
\item for purely noisy data, only local quartic and sextic couplings can be relevant to marginal in the large eigenvalue region (IR) domain. Moreover, there is a non-vanishing compact region around the Gaussian fixed point where all trajectories end toward the $\mathbb{Z}_2$ symmetric phase;
\item a strong enough signal makes the quartic and sextic local couplings irrelevant. Moreover, it induces a lack of symmetry restoration in the deep IR, for some trajectories which end continuously toward a broken phase. Hence, the strength of the signal plays the role of the inverse of the temperature $\beta = 1 / T$ in the familiar physics of phase transitions.
\end{itemize}

These results seem to link the presence of a signal to macroscopic properties of the associated field theory.
However, this approach focuses by construction on an equilibrium distribution, of maximum entropy.
In this paper, we propose to study the stability of this assumption regarding the presence of information in the spectrum, as the ratio signal/noise remains small enough. More precisely, we are aiming to establish a link between stability of the maximal entropy distribution and detection ability of the information merged in the noise. Note that, mathematically, what we mean by not detectable is the presence of sufficiently localized eigenvectors in the spectrum. 
\medskip

The model we consider in this paper takes the form of a dissipative Langevin equation reminiscent of the so-called \textit{model A}~\cite{livi2017nonequilibrium} encountered in out of equilibrium statistical physics, and describing evolution of a real field with respect to a time variable $t$, such that the equilibrium distribution for the field variable $\phi(t)$ corresponds to the maximum entropy estimator $\rho_{\text{eq}}[\phi]$. This would incidentally not be the first occurrence of a stochastic model in the analysis of correlations of large data systems.
In~\cite{chen2020searching}, for example, a Langevin dynamics is considered to study Hebbian learning~\cite{hebb2005organization} and patterns of synaptic activity of many connected biological neurons, and in particular to study the connection matrix at slow modes.
In statistical physics, specifically in spin glass theories, a dynamical point of view gives a complementary insight on the study of metastable states appearing at large in these types of systems.
The well-known Glauber model gives an example of such a dynamics for a system subjected to the influence of a thermal bath inducing random transitions~\cite{fischer1993spin}, and the so-called Model A is nothing but a coarse-grained
description of Glauber dynamics. More generally, the properties of time evolution and the role of fluctuations near the transition temperature is an important issue in the theory of phase transitions~\cite{hohenberg1977theory}. 
\medskip

In these examples, however, time is always an external variable.
But what about it in general?
Could we associate an abstract temporal dimension to a set of data, in order to get a universal framework of Langevin-like dynamics?
An interesting connection between temporality and statistics, known as the ``thermal time hypothesis'' in the literature, was established in the 1990s~\cite{rovelli1993statistical}.
The author points out the equivalence between the definition of a statistical equilibrium and the choice of a preferred time (i.e.\ a Hamiltonian flow), especially for covariant systems for which such a choice does not exist a priori.
From this point of view, a clock is nothing else than a system in equilibrium with the studied system, running linearly with the parameter of the Hamiltonian flow.
This point of view was generalized by Connes and Rovelli~\cite{rovelli1993statistical2}, who proved that, for quantum systems, there exists a one-parameter group of automorphisms (identified as a time stream), intrinsic to the von Neumann algebras, independently of the quantum state considered (this is a corollary of the Tomita-Takahashi theorem~\cite{connes1994neumann}).
Obviously, in these cases, there is a notion of temporality at the beginning.
For covariant systems, for example, the thermal time hypothesis only explains why, once a notion of equilibrium is fixed, a time stands out from the others and will tend to impose itself as a natural choice.

In our case, however, there is no notion of time at the beginning, and the question is: \emph{can we think of a canonical notion of time hidden behind the distributions of big data correlations?} 
We do not answer that fundamental question in this article.
We focus on a particular regime of non-equilibrium behaviour, such that the equilibrium distribution $\rho_{\text{eq}}[\phi]\sim e^{-\mathcal{S}[\phi]}$ is invariant to the notion of ``time'' introduced in the article.
Obviously, there is no single model of a stochastic equation that satisfies this condition, and we will keep things simple by considering a dissipative Langevin-type equation:
\begin{equation}
    \frac{\extd \phi}{\extd{t}}=-\frac{\partial \mathcal{S}}{\partial \phi(t)} + \eta(t),
    \label{SE}
\end{equation}
where $\eta(t)$ denote some Gaussian noise, with zero mean and delta-correlated $2$-points function $\langle \eta(t) \eta(t^\prime) \rangle \propto \delta(t - t^\prime)$.
Equation \eqref{SE} is probably one of the simplest to describe a dissipative process decreasing the energy.
It has been abundantly considered in the literature~\cite{DeDominicisbook,livi2017nonequilibrium}.
Obviously, a safer approach would be to follow the strategy described in~\cite{lahoche2021signal}, and deduce this equation from a particular problem, whose correlation spectrum tends towards a large-scale universal law.
This way, we could still use the same universality argument to justify this model, whatever the microscopic reality of the problem under study.
This is not the choice of this article, where we limit ourselves to motivating the choice of a stochastic approach, and reserve the underlying physical issues for a later work. 
\medskip 

Once this framework is established, we will focus on the study of its equilibrium dynamics for which we will be able to write a path integral, and build a RG \emph{à la Wilson} by partially integrating on the spectral degrees of freedom.
Note that it would also be possible to integrate jointly on the temporal degrees of freedom~\cite{Duclut_2017,lahoche2022functional,Lahoche:2021tyc,}.
However, we will not make this choice here, and the model presented in this article will be the subject of further studies.
We will use the Wetterich-Morris functional formalism, which is more adapted to the non-perturbative analyses required for this type of theory~\cite{Wett1,Wett2,Morris_19942,MORRIS_1994}.
\medskip

The main results of this article regarding the signal detection issue in nearly continuous spectra can be summed up as follows:
\begin{itemize}
    \item we assume, in the derivation of our effective field theory, that the system possesses non-translational invariance in time, and is in a large-time equilibrium dynamics;
    \item for a totally noisy spectrum, in the vicinity of MP's universality class, we observe that the system almost never returns to equilibrium and the flow diverges at finite scale, after a few RG steps (failure of the ergodic assumption);
    \item when the signal-to-noise ratio is low, the presence of a strong enough signal results in a sudden disappearance of divergences at least for a compact region of the phase space around the Gaussian fixed point, no longer violating the assumption of a return to equilibrium;
    \item the boundary between the two domains is marked by a critical value $\beta_c$ of the signal-to-noise ratio, which we are able to estimate. 
\end{itemize}
\medskip 

\paragraph{Outline} In section~\ref{sec1} we define the model and conventions, and provide technical preliminaries.
In section~\ref{sec2}, we provide a short presentation of the Wetterich-Morris formalism in this specific context, to study the stochastic field theory in an equilibrium dynamics phase.
We furthermore introduce the local potential approximation for analysing RG equation near phase transition.
In section~\ref{sec3}, we consider the previous formalism for spectra near the Marchenko-Pastur law, and we show that, universally, ergodicity is broken and equilibrium (almost) never reached for small enough signals (domain coarsening phase).
Furthermore, near the detection size (i.e.\ for the temperature $\beta$ near the critical value $\beta_c$) a slowing down effect is observed and correlation time becomes arbitrary large.
Finally, we conclude by enumerating some open questions that we plan to investigate in the future (section~\ref{sec4}).


\section{The model and associated path integral}\label{sec1}

This section provides the technical background underlying the study of this paper.
In the first part, we provide a short review of the framework and of the state of the art on the domain.
Next, we introduce the stochastic field theory model that we consider in the rest of this paper.
Furthermore, assuming to work in the equilibrium dynamics regime, we use the Martin-Siggia-Rose (MSR) formalism to write the dynamics as a path integral.
In the second part, we introduce the Wetterich-Morris equation and the local potential approximation (LPA) allowing investigating it in a non-perturbative regime.

\subsection{Technical preliminaries}\label{sec:technical_prelim}

As recalled in the introduction, standard PCA tools work well for spectra involving few discrete spikes isolated from the bulk.
In that case, only a very small number of eigenvalues is representative of a large fraction of the total variance, materialized by a gap in eigenvalues, for some $K = \Lambda$ in the fraction:
\begin{equation}
\zeta(K) \eqdef \frac{\sum_{\mu=0}^K \lambda_\mu}{\Tr \,C}\,,
\end{equation}
where $C$ denotes the \emph{covariance matrix} and $\{\lambda_\mu\}$ the set of its eigenvalues.
In practice, we focus on datasets defined by $N \times P$ matrices $X = \{ X_{ai} \}$, where indices $a$ and $i$ run along the sets $\{ 1, \cdots, P \}$ and $ \{ 1, \cdots, N \}$, respectively.
In other words, datasets are assumed to be large sets of big vectors.
Assuming the matrix $X$ is suitably mean-shifted, the covariance $C$ looks as a Wishart $N \times N$ matrix:
\begin{equation}
    C = \frac{X^T X}{P},
    \label{covarianceDEF}
\end{equation}
where $T$ means standard transposition.
Technically, since large variance contributions should dominate the spectrum, it is again convenient to work with the \emph{correlation matrix}, whose entries are defined as:
\begin{equation}
\tilde{C}_{ij} = \frac{C_{ij}}{\sqrt{C_{ii} C_{jj}}}\,.
\end{equation}
On the left of Figure~\ref{fig1}, we illustrate qualitatively the situation, where some discrete spikes capture a large fraction of the covariance matrix.
In~\cite{LahocheSignal2022}, we introduced the idea that the problem of distinguishing noisy from informational degrees of freedom for a nearly continuous spectrum near some random matrix universality class can be transposed to the RG study of some analogous field theory, describing an unconventional kind of matter filling an abstract space of dimension $1$.

As recalled in the introduction, it is interesting to notice that the definition of this field theory does not depend on the specific nature of data for which we are aiming to study the correlations.
Indeed, since we focus on the vicinity of some universal spectrum of random matrices, any effective or analogue field theory able to represent correlations for a specific problem can represent correlations for all datasets in the vicinity of the same universality class, if its mathematical formulation is general enough.
This is essentially the meaning of universality class, and the reflection of the fact that spectra are blind to the ``microscopic'' nature of degrees of freedom that they describe.
In this paper, we follow the working methodology of~\cite{LahocheSignal2022}.
As we aim to establish a link between properties of some effective field theory and signal detection, it is crucial in our numerical experiment to keep under control the spectra that we investigate, and in particular the signal-to-noise ratio threshold.
In practice, we have a parameter $T (\equiv \beta^{-1}) \in [0,1]$, such that, for $T=0$, $X$ becomes a purely i.i.d.\footnote{I.e., the entries of the matrix are independent and identically distributed variables.} random matrix of size $N \times P$, and the spectrum of $\tilde{C}(T=0)$ goes toward MP law weakly as $N,\, P \to \infty$, keeping $P / N \eqdef \alpha \geq 1$ fixed.
In formulas, denoting $x_i$ the eigenvalues of $\tilde{C}$ for $T=0$, we have~\cite{Potters1}:
\begin{equation}
\mu_{e}(x)
\eqdef
\frac{1}{N} \sum_{i=1}^N \delta(x-x_i)
\to
\mu_{MP}(x)
=
\frac{1}{2\pi\sigma^2}
\frac{\sqrt{(x-\lambda_-)(\lambda_+-x)}}{x \alpha} \textbf{1}_{[\lambda_-, \lambda_+]}\,,
\label{MP}
\end{equation}
where $\lambda_\pm = (1\pm \sqrt{\alpha} )^2$, $\sigma^2$ is the variance of the random entries and $\textbf{1}_{[\lambda_-, \lambda_+]}$ vanishes outside the domain $x \in [\lambda_-, \lambda_+]$.
In the continuum limit, we replace discrete sums by integrals:
\begin{equation}
\frac{1}{N} \sum_{i}f(x_i) \to \int \mu_{MP}(x)f(x)\, \extd{x}.
\end{equation}
In the rest of this paper we consider that such continuum approximation holds also for the experimental density spectra $\mu_{\text{exp}}(x)$ for $T \neq 1$, which we assume to be an implicit consequence of the nearly continuous approximation on which we focus in this paper.
Supposing to work in the vicinity of the MP law, the effective statistical field theory considered in~\cite{LahocheSignal2022} describes correlations for the scalar field $\phi(p) \in \mathbb{R}$ for $p\in \mathbb{R}$ playing the role of momenta.
The energy spectrum $p^2$ is assumed to take $N$ values, such that, in the continuum limit $N,\, P \to \infty$, they are distributed according to an (\emph{a priori} unknown) continuous bounded distribution $\rho(p^2)$.
The model is described by the probability density $\rho_{\text{eq}}[\phi]:=e^{-\mathcal{S}[\phi]}/Z$, the partition function being:
\begin{equation}
Z
=
\int [\extd{\phi}]\, e^{-\frac{1}{2} \sum_p \phi(-p)(p^2 + u_2)\phi(p) - U[\phi]}.
\end{equation}
In the expression, $[\extd{\phi}]$ denotes the path integral measure and the potential $U[\phi]$ corresponds to a ``local'' field theory~\cite{lahoche2020generalized}\footnote{%
    Strictly speaking, there is no locality principle underlying this theory because there is no ``background space''.
    We define locality from the formal similarity with standard field theory in Fourier space.
}
\begin{equation}
U[\phi]
=
N
\sum_{n=2}^K \frac{u_{2n}}{N^{n}}
\sum_{\{{p}_1,\cdots, {p}_{2n}\}} \delta\Big(\sum_{i=1}^{2n} {p}_i \Big) \prod_{i=1}^{2n} \phi({p}_i).
\end{equation}
As in standard field theory, we call the real constants $\{u_n\}$ the couplings, and $u_2$ the mass.
$N$ is the size of the correlation matrix $\tilde{C}(T)$, the dependency being crucial for the large $N$ limit to be well-defined.
In contrast with standard statistical or quantum field theories, the bare action $\mathcal{S}$ is essentially unknown, but the correlation functions of the theory are fixed ``experimentally'', using a data-driven approach (this is very similar to the approach of NN-QFT proposed in~\cite{erbin2021nonperturbative,erbin2022renormalization}).
Specifically, the $2$-point function $G(p^2)$,
\begin{equation}
G(p^2)
\eqdef
\frac{1}{Z} \int [\extd{\phi}] e^{-\mathcal{S}[\phi]}\phi(p)\phi(-p),
\end{equation}
which includes quantum corrections (Dyson's series) of the self-energy $\Sigma(p^2)$:
\begin{equation}
G(p^2)
=
\frac{1}{p^2+u_2}
+
\frac{1}{p^2+u_2} \Sigma(p^2) \frac{1}{p^2+u_2}
+
\cdots
=
\frac{1}{p^2+u_2-\Sigma(p^2)}\,,\label{quantumcorrectionspropa}
\end{equation}
is assumed to be equal, for each value of $p$, to an eigenvalue of the empirical correlation function $\tilde{C}(T)$.
For the Gaussian model, i.e.\ $u_{2n}=0\,\forall \, n>1$, the correspondence shows that $1 / (p^2 + u_2)$ should be identified with some $x_i$, and the mass $\mu_2$ corresponds to the inverse of the largest eigenvalue $\lambda_+$, assuming we translated the spectrum such that the smallest eigenvalue $\lambda_-$ is zero.
If the variable $x$ is distributed according to some distribution $\mu_0(x)$, we call $\rho_0(p^2)$ the induced distribution for $p^2$ in the Gaussian limit.
For $u_n \neq 0$, the propagator receives quantum corrections through the self energy $\Sigma(p^2)$, and in general $\rho(p^2) \neq \rho_0(p^2)$.
However, since we focus on the tail of the spectra, i.e.\ the region for $p^2 \ll 1$, we expect the derivative expansion and local potential approximation to work well.
In other words, we consider:
\begin{equation}
G(p^2) \approx \frac{1}{Zp^2+m^2},
\end{equation}
where $Z \eqdef 1 - \Sigma^\prime(0)$ and $m^2 \eqdef u_2-\Sigma(0)$.
In this approximation, we can assume that $p^2$ are distributed according to $\rho_0(p^2)$.
Furthermore, in the strict local potential approximation, $Z = 1$, as in~\cite{LahocheSignal2022}, we explicitly checked that this approximation indeed makes sense.
Hence, we postulate that all quantum corrections are included in the effective mass $m^2$, and:
\begin{equation}
\rho(p^2) \approx \rho_0(p^2).
\label{equalitydistribution}
\end{equation}
The properties of the local potential approximation for this model, and especially the relation between $\mathbb{Z}_2$ symmetry breaking and the strength of the signal were studied in~\cite{LahocheSignal2022} and references therein, and we recalled the main results in the introduction.
In this paper, we suggest another point of view: we consider dynamical rather than equilibrium phase transitions.
More specifically, we aim at studying the relationship between the presence of a detectable signal and the existence of equilibrium dynamics.
We recalled a first observation, a phase transition associated with the presence of a sufficiently strong signal, to which we can attach (at least formally) a critical temperature.
This has been investigated for an equilibrium theory, but we have not discussed the conditions for the existence of this equilibrium, or more precisely its stability.
However, out-of-equilibrium systems of this kind, associated with a system possessing different phases, exhibit a singular property known as \emph{coarsening} or \emph{phase ordering dynamics}.
If we move abruptly from a high-temperature (ergodic) regime to a temperature $T$ below the critical temperature $T_C$, the system is free to choose between several values of the order parameter, in different regions of the background space, and each region then evolves independently of the others, so that the system never returns to equilibrium.
The different phases ``grow'' as a function of time, with a scaling law $R(t)$, but the system remains self-similar at every instant~\cite{livi2017nonequilibrium}.
This scale invariance also seems in line with what would be expected for a totally noisy signal, and it would not be surprising if the effective field theory describing the large-scale collective behaviour of the degrees of freedom associated with such a signal exhibited such invariance.
We will see in the following that this is indeed the case for the effective kinetic theory model we will be proposing.
More precisely, we will see that a totally noisy signal in the vicinity of MP's universality class never allows a return to equilibrium, and the inverse of the critical temperature seems to cancel out at this limit $T_C^{-1} =0 $.
When this noisy spectrum is corrupted by a sufficiently strong signal (albeit within the signal/noise small enough limit), this critical temperature suddenly takes on a non-zero value.
Physically, it is tempting to interpret the existence of a region of the phase space supporting the equilibrium assumption as the manifestation of a macroscopic order, which in turn one would like to consider as relevant information.

\subsection{The model and Martin-Siggia-Rose formalism}

In the previous section, we motivated the analysis of a stochastic model describing a certain type of out of equilibrium process, suitably described by a Langevin equation of a particular type.
Because of the requirement that equilibrium theory corresponds to the field theory ``in equilibrium'' considered in reference~\cite{LahocheSignal2022}, we state the following ‘‘model A type" candidate~\cite{hohenberg1977theory} equation defined in Fourier space:
\begin{equation}
\boxed{
    \frac{\extd{~}}{\extd{t}} \varphi(p,t)
    =
    - (p^2+m^2) \varphi(p,t)
    - \frac{\partial U[\varphi]}{\partial \varphi(-p,t)}
    + \eta(p,t),
    }
\label{Langevin1}
\end{equation}
which describes the temporal evolution of the random field $\varphi(p,t)$, while keeping the notation $\phi\equiv \phi(p)$ for the equilibrium field variable.
The white noise $\eta(p,t)$ is assumed to be Gaussian, with zero mean and variance:
\begin{equation}
\langle \eta(p,t) \eta(p^\prime,t^\prime) \rangle
=
2~T~\delta_{p,-p^\prime} \delta(t-t^\prime),
\end{equation}
where the notation $\langle X \rangle$ is the average over the normalized noise distribution $\extd{\mu}(\eta)$:
\begin{equation}
\langle X \rangle
\eqdef
\int \extd{\mu}(\eta)\,  X.
\end{equation}
The parameter $T$ identifies the temperature of equilibrium states, and we set $T=1$ in the rest of this paper (see ~\cite{ZinnJustinBook2}).
The stochastic process described by the equation \eqref{Langevin1} can be equivalently expressed in terms of the probability density $P(\varphi, t)[\extd{\varphi}]$ for the field to be in the functional domain $[\varphi, \varphi+[\extd{\varphi}]]$, starting from some initial condition at $t = 0$.
This density probability reads explicitly as:
\begin{equation}
P(\varphi, t)
\eqdef
\left\langle \prod_p\delta\left(\varphi_\eta(p,t)-\varphi(p)\right) \right\rangle,
\end{equation}
where $\varphi_\eta(p,t)$ is a formal solution of equation \eqref{Langevin1}, for a given $\eta$.\footnote{%
    Note that the solution is unique as soon as the initial conditions are fixed because the equation is first order with respect to time.
The probability density is obviously normalized to $1$, thanks to the normalization of $\extd{\mu}(\eta)$.
    }
Furthermore, it obeys a Fokker-Planck equation~\cite{ZinnJustinBook2}, whose stationary solutions are\footnote{%
    For $T \neq 1$, the equation is replaced by $P_{\text{eq}}[\phi]\sim e^{-\mathcal{S}/T}$, and we indeed identify $T$ as the temperature for equilibrium states, as soon as $\mathcal{S}$ is identified with the Hamiltonian.
}
\begin{equation}
P_{\text{eq}}[\phi]
\equiv \rho_{\text{eq}}[\phi]
\propto
\exp \left(-\frac{1}{2} \sum_p \phi(-p) (p^2+m^2) \phi(p)-U[\phi]\right)
\equiv
\exp (-\mathcal{S}[\phi])\,,
\end{equation}
which is also the expected probability density for late time.
Indeed, introducing the ``wave function''\footnote{
    The terminology makes sense in imaginary time.
} $\Psi \eqdef e^{\mathcal{S}/2}P$, we have $\dot{\Psi}=-\hat{H}\Phi$, with
\begin{equation}
\hat{H}
\eqdef
\frac{1}{2}
\left(-\frac{\partial}{\partial \varphi} + \frac{1}{2}\frac{\partial \mathcal{S}}{\partial \varphi}\right)
\left(\frac{\partial}{\partial \varphi}+ \frac{1}{2}\frac{\partial \mathcal{S}}{\partial \varphi}\right).
\end{equation}
The fundamental state is such that $\hat{H}\Psi_0=0$, namely $\Psi_0 = e^{-\mathcal{S}/2}$, which exists if and only if $\Psi_0$ is normalizable,
\begin{equation}
\langle \Psi_0 \vert \Psi_0 \rangle = \int [\extd{\phi}]\, e^{-\mathcal{S}[\phi]}.
\end{equation}
In that case, we expect the system to return towards equilibrium:
\begin{equation}
\lim\limits_{t \to +\infty} P(\phi,t) = P_{\text{eq}}[\phi].
\end{equation}
In the equilibrium dynamics regime, the Martin-Siggia-Rose (MSR)~\cite{martin1973statistical,DeDominicisbook} formalism allows representing information about correlations of the field at different time as a partition function defined by a path integral over two fields.
This partition function $\mathcal{Z}[J]$ can be defined as follows:
\begin{equation}
\mathcal{Z}[J]
\eqdef
\left\langle e^{\int \extd{t} \sum_p J(-p,t) \varphi(p,t)} \right\rangle,
\label{averageZ}
\end{equation}
which, according to the MSR strategy, can be rewritten as:
\begin{equation}
\mathcal{Z}[J, \tilde{J}]
=
\int [\extd{\varphi]} [\extd{\tilde{\varphi}}]\,
e^{-\mathcal{S}[\varphi,\tilde{\varphi}] + 
\int \extd{t} \sum_p J(-p,t) \varphi(p,t) + 
\int \extd{t} \sum_p  \tilde{J}(-p,t) \tilde{\varphi}(p,t)}\,,\label{MSRZ} 
\end{equation}
where we included a source term $\tilde{J}$ for the auxiliary field $\tilde{\varphi}$ as well.
The MSR action is defined as:
\begin{equation}
\mathcal{S}[\varphi,\tilde{\varphi}]
\eqdef
\int \extd{t}
\sum_p
\left(
    \frac{\tilde{\varphi}(-p,t)\tilde{\varphi}(p,t)}{2}
    +
    \tilde{\varphi}(-p,t)
    \left(
        i\dot{\varphi}(p,t) + \frac{1}{2} \frac{\partial \mathcal{S}}{\partial \varphi(p,t)}
    \right)
\right).
\end{equation}
Notice that we use the Itô prescription for the computation of path integral, imposing in particular $\theta(0) = 0$ for the Heaviside theta function involved in the computation of the average \eqref{averageZ}, see~\cite{ZinnJustinBook2}.
In the literature, the auxiliary field is known as a \emph{response field}, for reasons that we do not explain here, though the interested reader may consult~\cite{Aron_2010}.


\section{Functional renormalization group}\label{sec2}

There are many incarnations of the original Wilson idea of renormalization group (RG), but many of them are not suitable for non-perturbative analysis, like the Polchinski equation, even if they are formally accurate.
Nowadays, the most powerful approach for non-perturbative aspects of RG is the Wetterich-Morris formalism, that we will use in this paper --- for more details on this topic, the reader may consult the standard reference~\cite{Delamotte1}.
In the first subsection, we introduce this general formalism for the field theory that we constructed in the previous section.
Let us mention right away that, in our case, the Wetterich-Morris equation cannot be solved exactly (as it is often the case in physics\footnote{%
    This is essential.
    It is often pointed out that the solutions proposed in this framework are approximate, which is however a characteristic of physics, not of the Wetterich formalism.
}), and approximations will be necessary.
We will describe them in the second subsection.
More details on the functional RG in this context can be found in the recent references~\cite{lahoche2022functional,Lahoche:2021tyc} of the same authors or the standard references~\cite{Canet_2011,Duclut_2017,Dupuis_2021}.
Some recent papers cover also the topic of RG in and out of equilibrium stochastic process, see for instance~\cite{wilkins2021functional,wilkins2021functional2}.

\subsection{Functional renormalization in a nutshell}\label{sec121}

\begin{figure}
    \centering
    \includegraphics[width=0.7\linewidth]{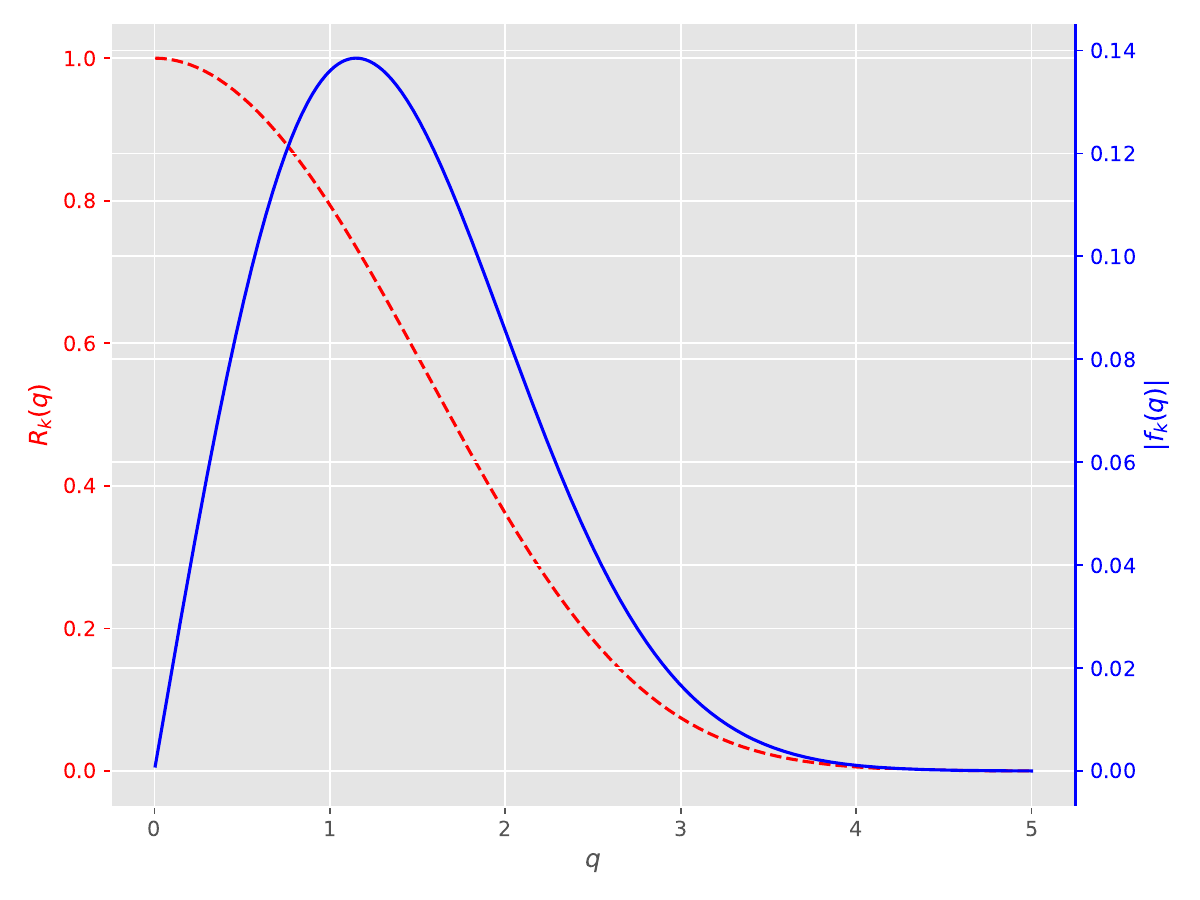}
    \caption{Behaviour of the threshold function for a typical regulator $R_k$ (the dashed blue curve).}
    \label{figRegul}
\end{figure}

Let us consider a field theory whose partition function is given by the path integral:
\begin{equation}
Z = \int [\extd{\phi}]\, e^{-\mathcal{S}[\phi]}\,.
\end{equation}
In the Wilsonian point of view for RG, microscopic degrees of freedom are integrated out.
The fundamental cut-off is rescaled at each step, to provide $\beta$-functions that describe how partial integration changes the couplings involved in the classical action.
In the point of view proposed by Wetterich and Morris~\cite{Delamotte1}, however, the fundamental scale remains fixed, but a running parameter playing the role of an infrared cut-off suppresses large-scale contributions in the effective action of ultraviolet modes.
This infrared cut-off is denoted by $k$, and we modify the classical action as
\begin{equation}
\mathcal{S}[\phi] \to \mathcal{S}[\phi] + \Delta\mathcal{S}_k[\phi],
\label{modifiedaction}
\end{equation}
where the \emph{regulator} $\Delta\mathcal{S}_k[\phi]$, which is of degree $2$ in the field $\phi$ looks formally as a momentum and scale dependent mass term, designed to decouple long-range energy modes (with respect to the scale $k$).
Explicitly, for a field theory in $D$ dimensions ($x\in \mathbb{R}^D$, $\phi:\mathbb{R}^D \to \mathbb{R}$):
\begin{equation}
\Delta \mathcal{S}_k[\phi]
\eqdef
\frac{1}{2} \int \mathrm{d}^D x \mathrm{d}^D y\, \phi(x) R_k(x-y) \phi(y).
\end{equation}
A typical shape for (the Fourier transform of) the function $R_k(x-y)$ is shown in Figure~\ref{figRegul}.
In the figure, we show as well the behaviour of the threshold function:
\begin{equation}
f_k(q^2) = \frac{\partial_q R_k(q^2)}{q^2 + R_k(q^2)}
\end{equation}
where $q$ denotes the momentum in Fourier space.
The function $f_k$ is the typical integrand involved in the flow equation, and we see that the regulator reduces the windows on momenta to a small domain around $q^2\approx k^2$.
In practice, the regulator shape is designed such that:
\begin{itemize}
    \item $R_{k}(q^2)\to 0$ as $\vert q \vert \to 0$, $k\to 0$, meaning that all the modes are integrated out in the deep infrared limit. 
    \item $R_{k}(q^2)\to k^2$ as $\vert q \vert/k \to 0$, meaning that small energy modes decouple from long-distance physics.
    \item $R_{k}(q^2)\simeq 0$ as $\vert q \vert/k \gg 1$, meaning that high-energy modes remain unaffected by the regulator and are integrated out. 
\end{itemize}
The Wetterich approach focuses on the effective average action $\Gamma_k$, which can be defined as the effective action for the ultraviolet modes (i.e.\ large with respect to the infrared scale $k$).
This effective action is defined as the slightly modified Legendre transform of the free energy $W_k$:
\begin{equation}
\Gamma_k[M] + \Delta\mathcal{S}_k[M] = \int \mathrm{d}^D x\, J(x) M(x) - W_k[J],
\end{equation}
where $J(x)$ is the source field, $M(x)$ is the classical field:
\begin{equation}
M(x)
\eqdef
\frac{\delta W_k}{\delta J(x)},
\end{equation}
and the free energy $W_k$ is defined as:
\begin{equation}
e^{W_k[J]}
\eqdef
\int [\extd{\phi}]\, e^{-\mathcal{S}[\phi] - \Delta\mathcal{S}_k[\phi] + \int \mathrm{d}^D x\, \phi(x) J(x)}.
\end{equation}

Given the properties of the regulator, the effective average action $\Gamma_k$ is a smooth interpolation between the fundamental classical action $\mathcal{S}$ and the full effective action $\Gamma$ --- that is, the true Legendre transform of the free energy when the regulator function is removed.
For some ultraviolet cut-off $\Lambda \gg 1$, we thus have:
\begin{enumerate}
    \item $\Gamma_{k=\Lambda} \simeq \mathcal{S}$ because $R_{k\to \Lambda} \sim \Lambda^2$ and quantum fluctuations are almost frozen;
    \item $\Gamma_{k=0} \equiv \Gamma$ because $R_{k=0} = 0$ and all the fluctuations are integrated out.
\end{enumerate}
The equations describing the variation of the interpolation $\Gamma_k$ is the Wetterich-Morris equation, which explicitly reads as:
\begin{equation}
k \frac{\mathrm{d}}{\mathrm{d}k}\Gamma_k
=
\frac{1}{2} \int \mathrm{d}^D x \mathrm{d}^D y\, \left(k \frac{\mathrm{d}}{\mathrm{d}k} R_k(x-y) \right) G_k(x-y),
\label{Wett}
\end{equation}
where the $2$-point function $G_k$ is the formal inverse of the $1$PI $2$-point function.
In momenta:
\begin{equation}
G^{-1}(q^2) = \Gamma_k^{(2)}(q^2)+R_k(q^2),
\label{G}
\end{equation}
where $\Gamma_k^{(2n)}$ denotes the $2n$-th order functional derivative of $\Gamma_k$ with respect to the classical field $M$.
The flow equation \eqref{Wett} is \emph{exact}, but cannot be solved exactly in general, and approximations are required.
Our aim in this paper is to study the RG corresponding to the MSR generating functional \eqref{MSRZ}.
The method is described in recent works~\cite{Duclut_2017,Lahoche:2021tyc,lahoche2022functional}, and the reader may consult them and references therein for more details.
The regulator has to be of the form:
\begin{equation}
\begin{split}
\Delta \mathcal{S}_k
\eqdef
& \int \extd{t} \extd{t}^\prime \sum_{p} r_k(p^2) \times
\\
& \Big(i \tilde{\varphi}(-p,t) \rho_k^{(1)}(t-t^\prime)  \varphi(p,t^\prime)+\frac{1}{2}\tilde{\varphi}(-p,t)\rho_k^{(2)}(t-t^\prime)\tilde{\varphi}(p,t^\prime)
\Big)\,,
\end{split}
\label{regulatordef}
\end{equation}
where $r_k(p^2)$ is the Litim regulator~\cite{Litim_2001} over the eigenvalues:
\begin{equation}
r_k(p^2)
\eqdef
(k^2-p^2) \theta(k^2-p^2)\,.
\end{equation}

Time reversal symmetry of the MSR classical action corresponds to the field transformation:
\begin{equation}
\varphi^\prime(p,t)
\eqdef
\varphi(p,-t),
\qquad
\tilde{\varphi}^\prime(p,t)
\eqdef
\tilde{\varphi}(p,-t)+2i\dot{\varphi}(p,-t).
\label{fieldtransform}
\end{equation}
The requirement that the modified action \eqref{modifiedaction} remains invariant under time reversal for all $k$ implies:
\begin{equation}
\rho_k^{(1)}(t)-\rho_k^{(1)}(-t)+\dot{\rho}_k^{(2)}(-t)-\dot{\rho}_k^{(2)}(t)=0.
\label{relreg}
\end{equation}

In this paper we do not consider a coarse-graining over time (i.e.\ over frequencies), and we set $\rho_k^{(1)}(t)=1\Rightarrow \rho_k^{(2)}(t)\equiv 0$ because of \eqref{relreg}.
For a comparison between $\dot{\rho}_k^{(1)}(t)=0$ v.s. $\dot{\rho}_k^{(1)}(t)\neq 0$ for a model close to the one considered in this paper, the reader may consult~\cite{lahoche2022functional}.
We expect that our approximation is as accurate as necessary for a proof of concept.
In the next section, we discuss the local potential approximation (LPA). 

\subsection{Local potential approximation}\label{sec122}

In~\cite{lahoche2021signal}, we considered the LPA for the equilibrium theory.
In this section, we introduce the same formalism for the out of equilibrium field theory considered in this paper.
Once again, more details about computations can be found in~\cite{lahoche2022functional} of the same authors.
In this approximation, we assume that the classical field has a macroscopic value $\chi$ for the deep IR component of the classical field $M(p)$ corresponding to the average value for the field $\varphi$:
\begin{equation}
M(p)=\sqrt{N\chi} \, \delta_{0,p}.
\label{IRfield}
\end{equation}
We furthermore assume the following ansatz for $\Gamma_k$:
\begin{align}
\Gamma_k[\Xi]
\eqdef
\Gamma_{k,\text{kin}}[\Xi]
+
\frac{1}{2}N\int \extd{t} \sum_p\,i \tilde{\varpi}(-p,t)\frac{\partial U_k}{\partial M(p,t)},
\label{ansatztruncation}
\end{align}
where $\Xi=(\tilde{\varpi},M)$ denotes collectively the classical fields, i.e.\ the means values respectively for the response field $\tilde{\varphi}$ and the random field $\varphi$, and:
\begin{equation}
\begin{split}
\Gamma_{k,\text{kin}}[\Xi]
\eqdef
\int \extd{t} \sum_p\, \Bigg( & Y_k \frac{\tilde{\varpi}(-p,t)\tilde{\varpi}(p,t)}{2} 
\\
&\qquad  +i \tilde{\varpi}(-p,t) \bigg(Y_k \dot{M}(p,t)+Z_kp^2 M(p,t)\bigg)\Bigg)
\end{split}
\end{equation}

Since we focus on strict LPA, we furthermore assume (see~\cite{Delamotte1}):
\begin{equation}
Y_k=Z_k=1\,,\quad \forall\, k.
\end{equation}
Furthermore, the equilibrium average action is ($m(p)$ denotes here the equilibrium classical field):
\begin{equation}
\begin{split}
\Gamma_{k,\text{eq}}[m] &
\equiv \frac{1}{2}\sum_p m(-p)p^2m(p)+\frac{1}{2} U_k[m^2] \\
& =\frac{1}{2}\sum_p m(-p)(p^2+\mu^2_{\text{eq}}(k))m(p)+\mathcal{O}(m^4),
\end{split}
\end{equation}
and the physical mass $\mu^2_{\text{eq}}(k=0)$ is identified with the inverse of the largest eigenvalue of the empirical density spectra $\mu_{\text{exp}}(\beta)$~\cite{LahocheSignal2022}\footnote{
    This is indeed a strong but crucial simplification of the LPA to store all the quantum fluctuations effects for the $2$-point function into the effective mass.
}.
Furthermore, in the same reference, the authors checked for the equilibrium theory in particular that wave function renormalization is indeed a small correction, and we expect this approximation to make sense for the purpose of this paper.
For the classical configuration \eqref{IRfield}, we have:
\begin{equation}
\frac{1}{2}\frac{\partial U_k}{\partial M(p)}\to U_k^\prime[M^2] M(p), 
\end{equation}
and $U_k[M^2]$ for uniform macroscopic field is expected to admit a power expansion around some (running) minimum $\kappa(k)$:
\begin{equation}
U_k(M^2) = \frac{\mu_2(k)}{2}\left(\frac{M^2}{N}-\kappa\right)^2+\frac{\mu_3(k)}{3}\left(\frac{M^2}{N}-\kappa\right)^3+\cdots
\label{explicitUk}
\end{equation}

The flow equation for the effective potential $U_k^\prime$ can be derived from the Wetterich equation \eqref{Wett}, imposing condition \eqref{IRfield} on both sides of the equation.
From the truncation, we get formally:
\begin{equation}
\int \extd{t}\, N \frac{\mathrm{d}}{\extd{s}} U_k^\prime[N\chi]
=
-i \frac{1}{\sqrt{N\chi}} \int \extd{t}
\sum_p\, \frac{\mathrm{d}}{\extd{s}} \frac{\delta \Gamma_k}{\delta \tilde{\varpi}(p,t)}
\equiv
-\frac{1}{2}\,\vcenter{\hbox{\includegraphics[scale=1]{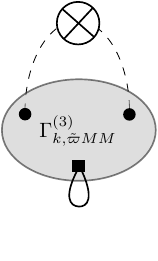}}}
\,,
\label{floweqpotential1}
\end{equation}
where $\extd{s} \eqdef \extd{(\ln k)}$.
The last diagram corresponds to the right hand side of the flow equation \eqref{Wett}: the black dots and black squares represent respectively the fields $M$ and $\tilde{\varpi}$, the solid self-loop is the sum and time integral over the variable of $\tilde{\varpi}(p, t)$, the dotted edge materializes the effective propagator $G_k$ (equation \eqref{G}), and the crossed circle represents the regulator.
The $3$-points function $\Gamma_k^{(3)}$ can be easily computed:
\begin{equation}
\begin{split}
\frac{\delta^3\Gamma_k}{\delta \tilde{\varpi}(p,t) \delta M(p_1,t_1)\delta M(p_2,t_2)}
=
& N
\Big(
    2{U}_k^{\prime\prime}
    \Big[\delta_{p,p_1} \delta_{0,p_2}+\delta_{p,p_2} \delta_{0,p_1}+\delta_{p_1,p_2} \delta_{0,p}\Big]
\\
&
    +4 \delta_{0,p}\delta_{0,p_1}\delta_{0,p_2} \chi N\,{U}_k^{\prime\prime\prime}
\Big)
\\
&
\times \sqrt{N\chi}\delta(t-t_1)\delta(t-t_2).
\label{decomp3pts2}
\end{split}
\end{equation}
It is convenient to introduce the potential $\mathcal{U}_k$ such that:
\begin{equation}
\mathcal{U}_k[\chi] \eqdef U_k[M^2=N\chi].
\end{equation}
Hence, working in Fourier space, and because of the integral:\footnote{%
    It comes directly from residue theorem, assuming the function $f(x)$ does not vanishes.
}
\begin{equation}
\int_{-\infty}^{+\infty} \frac{\extd{\omega}}{2\pi} \frac{1}{(i\omega+f(x))^2(-i\omega+f(x))}
=
\frac{1}{4}\frac{1}{(f(x))^2},
\end{equation}
the flow equation for $\mathcal{U}_k^\prime[\chi]$ reads explicitly:
\begin{equation}
\frac{\mathrm{d}}{\extd{s}} \mathcal{U}_k^\prime[\chi]
=
-2\frac{3{\mathcal{U}}_k^{\prime\prime}[\chi]
+2\chi {\mathcal{U}}_k^{\prime\prime\prime}[\chi]}{(k^2+\mu^2)^2} k^2 \intpk,
\label{flowUk}
\end{equation}
where we used the explicit expression for the derivative of the Litim regulator:
\begin{equation}
\frac{\mathrm{d}}{\extd{s}}r_k(p^2) = 2k^2\theta(k^2-p^2),
\end{equation}
and the effective mass $\mu^2$ is:
\begin{equation}
\mu^2 \eqdef \mathcal{U}_k^\prime[\chi]+2\chi \mathcal{U}_k^{\prime\prime}[\chi].
\end{equation}
Furthermore, we replaced the discrete sum over $p$ by an integral, taking the continuous limit for the density spectrum:
\begin{equation}
\frac{1}{N}\sum_p \theta(k^2-p^2) \to 4 \intpk.
\end{equation}

\subsection{Scaling and dimensions}

For a power-law distribution $\rho(p^2) \sim (p^2)^{\delta}$, as it is the case for standard field theory,\footnote{For a field theory over $\mathbb{R}^d$, we have $\rho(p^2)\sim (p^2)^{\frac{d-2}{2}}$.} the loop integral in \eqref{flowUk} behaves as $k^{2\delta+2}$.
The explicit dependency on $k$ can be cancelled on both sides of the flow equation, working with dimensionless quantities.
Specifically, this allows taking into account the rescaling of the fundamental scale after each partial integration, to keep the large-scale physics unchanged.
In our case, however, the distribution is not a power law, and the integral in \eqref{flowUk} does not behave as such.
It that case, we cannot discard the explicit dependency on $k$.
As pointed out in~\cite{lahoche2021signal}, the best compromise is to move this dependency to the level of the linear term, that we usually call \textit{canonical dimension} in the RG literature.
Hence, according to~\cite{lahoche2021signal}, we introduce the new parameter $\tau$, defined such that:
\begin{equation}
\extd{\tau}
\eqdef
\extd{\left[\ln \intpk\right]}.
\label{dtau}
\end{equation}
For a power law distribution obviously, $\extd{\tau} \propto \extd{\ln k}$.
Since the eigenvalues $p^2$ have to scale as $\mu^2$, it is convenient to introduce the \emph{dimensionless effective mass}:
\begin{equation}
\bar{\mu}^2 \eqdef k^{-2} \mu^2.
\end{equation} 
We introduce the notation:
\begin{equation}
\dot{X} \eqdef \frac{\extd{X}}{\extd{\tau}}.
\end{equation}
for the $\tau$ derivative, and we denote as $\dim_\tau(X)$ the canonical dimension (i.e.\ the opposite of the linear term in the flow equation for $X$).
Since the flow equation for $\mu^2$ has to be multiplied by $\dot{s}$ to express the flow in the $\tau$ variable, the canonical dimension for $\mu^2$ is:
\begin{equation}
\dim_\tau(\mu^2)
\eqdef th
\dim_\tau(\mathcal{U}_k^\prime) = 2\dot{s}.
\label{dimmass}
\end{equation}
Henceforth, we define as $\bar{X}$ the dimensionless version of the quantity $X$.
Using the $\tau$ parameter, the flow equation for $U_k^\prime$ \eqref{flowUk} becomes:
\begin{equation}
\dot{\mathcal{U}}_k^\prime[\chi]
=
- 2\, k^2\, \frac{3{\mathcal{U}}_k^{\prime\prime}[\chi]
+ 2\chi {\mathcal{U}}_k^{\prime\prime\prime}[\chi]}{(1+\bar{\mu}^2)^2}
\left(\frac{\rho(k^2)}{k^2} \dot{s}^2\right).
\label{flowUktau}
\end{equation}
To derive the flow equations for dimensionless couplings, it is convenient to work with a flow equation with fixed $\bar{\chi}$.
The flow equation \eqref{flowUk} is, however, written at a fixed $\chi$.
To convert one into the other, let us observe that:
\begin{equation}
\dot{\mathcal{U}}_k^\prime[\chi]
=
k^2
\bigg[
    \dot{{\bar{\mathcal{U}}}}_k^{\prime}[\bar{\chi}]
    +
    \dim_\tau(\mathcal{U}_k^\prime)\, \bar{\mathcal{U}}_k^\prime[\bar{\chi}]
    -
    \dim_\tau(\chi)\, \bar{\chi}\, \bar{\mathcal{U}}_k^{\prime\prime}[\bar{\chi}]
\bigg]. 
\label{eqLagrangebis}
\end{equation}
The dimension of $\dim_\tau(\mathcal{U}_k^\prime)$ has been computed in equation \eqref{dimmass}.
From the flow equation \eqref{flowUktau}, we define:
\begin{equation}
{\bar{\mathcal{U}}}_k^{\prime\prime}
\eqdef
{\mathcal{U}}_k^{\prime\prime}\frac{\rho(k^2)}{k^2} (\dot{s})^2.
\label{dimquartic}
\end{equation}
Obviously $\chi {\mathcal{U}}_k^{\prime\prime\prime}$ and $\mathcal{U}_k^{\prime\prime}$ must have the same dimension, and the dimension of the field $\chi$ can be computed from \eqref{dimquartic} and \eqref{dimmass}:
\begin{equation}
\chi \eqdef \left(\rho(k^2) \dot{s}^2\right)\bar{\chi},
\end{equation}
leading to:
\begin{equation}
\dim_\tau(\chi)
=
\dot{s} \frac{d}{ds} \ln \left(\rho(k^2) \dot{s}^2\right).
\end{equation}
Finally, the flow equation for the dimensionless potential reads explicitly:
\begin{equation}
\boxed{
    \dot{\bar{\mathcal{U}}}_k^\prime[\bar\chi]
    =
    -\dim_\tau(\mathcal{U}_k^\prime) \bar{\mathcal{U}}_k^\prime[\bar{\chi}]
    +\dim_\tau(\chi) \bar{\chi} \bar{\mathcal{U}}_k^{\prime\prime}[\bar{\chi}]
    -2\frac{3\bar{\mathcal{U}}_k^{\prime\prime}[\bar{\chi}] + 2\bar{\chi}
    \bar{{\mathcal{U}}}_k^{\prime\prime\prime}[\bar{\chi}]}{(1+\bar{\mu}^2)^2}.
}\label{flowequationfull}
\end{equation}
This equation provides, in the LPA approximation, the full behaviour of the RG.
The formalism being introduced, we will study in the next sections the RG flow for slightly deformed spectra around MP law. 

\begin{remark}
According to the discussion in section~\ref{sec:technical_prelim}, in equation \eqref{equalitydistribution}, the distribution $\rho(p^2)$ is assumed to be equal to the distribution $\rho_0(p^2)$ for the Gaussian theory. 
\end{remark}

\section{Numerical flow analysis and the signal track}
\label{sec3}

This section aims at exploring the behaviour of the RG for the theory described in the previous section (equation \eqref{flowequationfull}).
We start by studying the case of a totally noisy spectrum, that is, the analytical MP distribution.
We then investigate the qualitative changes in the behaviour of the RG as a signal disturbs the noise matrix.

\subsection{Numerical methodology}

In order to simulate the flow equation \eqref{flowequationfull}, we use the Python package \texttt{py-pde}~\cite{Zwicker:PyPDE:2020}.
It enables the efficient simulation of partial differential equations of the general form
\begin{equation}
    \dot{u}(\mathbf{x}, t)
    =
    \mathcal{D}[u(\mathbf{x}, t)] + \eta(u, \mathbf{x}, t),
\end{equation}
where $\mathcal{D}$ is a (generically non-linear) differential operator, $\dot{u}$ is the usual ``time'' derivative, and $\eta$ is a noise term (in our case $\eta \equiv 0$).

\begin{figure}[t]
    \centering
    \includegraphics[width=0.7\linewidth]{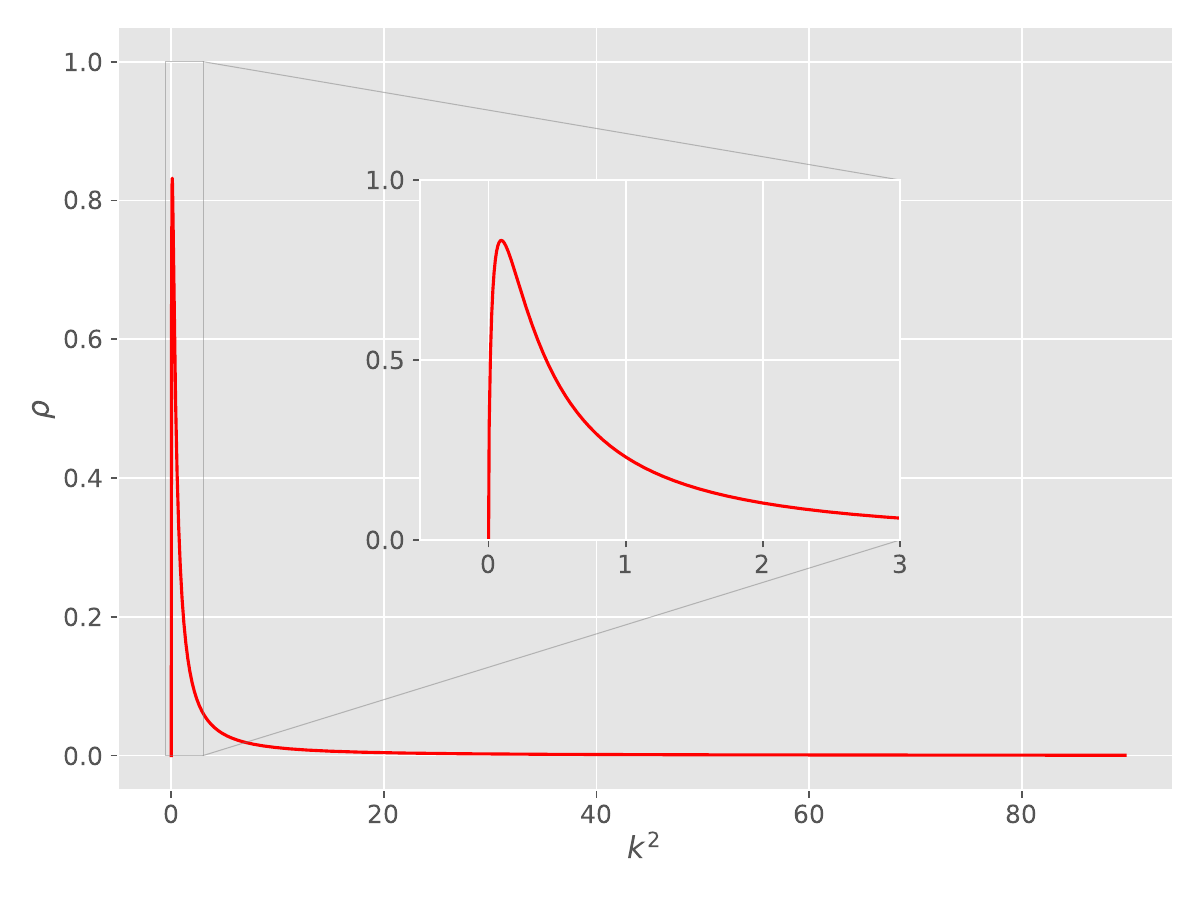}
    \caption{%
        Typical shape of $\rho(p^2)$ for MP law.
        Note that we labelled the abscissa variables with $k^2$, the renormalization group scale.
    }
    \label{inverse_MP}
\end{figure}

As we are interested in the universality class of the MP distribution, we first focus on its analytic form \eqref{MP}.
We inverse it in order to obtain the functional form of the momenta $\rho(p^2)$ of the field theory (see Figure \ref{inverse_MP}):
\begin{equation}
    \rho(p^2) = \mu_{MP}\left( \frac{1}{p} \right)\, \frac{1}{p^2}.
\end{equation}
We successively translate it such that $\rho(0) = 0$.
We then consider several empirical realisations to follow the deformation from the theoretical distribution.
We consider a noisy signal given by the definition \eqref{covarianceDEF}, where the entries of the matrix $X$ are independent and identically distributed (i.i.d.), with zero mean and unit variance.
In the large $N,\, P \to \infty$ limit, but such that $P/N \eqdef \alpha < \infty$, the spectrum of $C$ approximates the MP distribution defined by equation \eqref{MP} -- see also Figure~\ref{fig2}.
To deform the spectrum of the universal class, we build a fixed-rank matrix $S$, added on top of the purely random matrix $X$.
Specifically, we build an empirical matrix $Y \in \mathbb{R}^{N \times P}$ such that\footnote{%
    Notice the slight change of notation w.r.t.\ Section \ref{sec:technical_prelim}, in order to conform to the standard literature in data analysis.
    All definitions can be recovered by either transposing the matrix $X$ in the previous sections, or by inversing $P \longleftrightarrow N$.
}:
\begin{equation}
    Y = Z + \beta\, S,
\end{equation}
where $Y = (y_{ij})_{i \in [1, N],\, j \in [1, P]}$ for $y_{ij} \sim \mathcal{N}(0, 1)$, and $\beta \in [0,\, 0.5]$ in these simulations.
The signal matrix $S$ represents the added signal.
In the investigations carried out in references~\cite{lahoche2021signal,LahocheSignal2022}, the authors considered the spectrum of a correctly normalized image to materialize this signal.
Here, we adopt a different strategy, which allows us to maintain even greater control over the experimental parameters.
We build two rectangular matrices
\begin{equation}
    U = (u_{ir})_{i \in [1, N],\, r \in [1, R]}
    \qquad \text{and} \qquad
    V = (v_{rj})_{r \in [1, R],\, j \in [1, P]}
\end{equation}
for a fixed choice of $R < N,\, P$, such that $u_{ir} \sim \mathcal{N}(0, 1)$ and $v_{rj} \sim \mathcal{N}(0, 1)$.
We then consider the product
\begin{equation}
    S \eqdef \frac{U\, V}{\sqrt{R}}
\end{equation}
as the definition of the signal matrix.
The covariance matrix has the usual definition (row-wise variables):
\begin{equation}
    C \eqdef \frac{Y^T\, Y}{N}.
\end{equation}
We rely on the interpolation of the (inverse) spectrum of the eigenvalues of $C$ for the subsequent steps.
Specifically, we fit a smoothed 2nd degree polynomial to the density histogram of the eigenvalues / momenta, using a B-spline representation of the curve.
Integrations were carried out using the quadrature technique, while differentiation uses central differences to minimize the error.

For our numerical investigations, we set $N = 10^4$, $P = 8 \times 10^3$, and $R = 2500$.
Moreover, we set the general initial condition
\begin{equation}
\bar{\mathcal{U}}_{k=0}[\bar{\chi}]
=
\bar{\mu}_1 \bar{\chi}
+
\frac{1}{2} \, \bar{\mu}_2 \bar{\chi}^2
+
\frac{1}{3} \, \bar{\mu}_3 \bar{\chi}^3
+
\frac{1}{4} \, \bar{\mu}_4 \bar{\chi}^4
\label{eq:pot_initial}
\end{equation}
on the effective potential.

The simulation of the dynamic equation \eqref{flowequationfull} has been carried out in a frequency domain centred on the low momentum scale of the underlying MP distribution.
That is, we chose $k \in [k_{-},\, k_{+}]$ in $\tau = \tau(k)$ (see \eqref{dtau}) such that:
\begin{equation}
    k_{\pm}
    =
    \frac{1}{\lambda_+ \pm a},
\end{equation}
where $\lambda_+$ is defined in \eqref{MP}, and $a = 0.35$ in the simulations presented in this article.
We set the simulation on a grid of $10^3$ points in the interval $\bar{\chi} \in [0,\, 1]$.

At least as far as the $2$-point function is concerned, the considered field theory is defined in the IR by construction (we have the exact function, including all quantum corrections –- see \eqref{quantumcorrectionspropa}).
This contrasts with the usual situation in field theory, where we define a microscopic theory whose large-scale effects are studied by the RG; here, our definition of microscopic theory is based solely on the approximation (LPA) used to describe the RG flow.
For this reason, the microscopic theory should not be considered as realistic, in the sense that it does not reflect the reality of the microscopic degrees of freedom associated with the spectrum under consideration.
It is only a projection of the actual theory, describing at a large distance the collective behaviour of these degrees of freedom. Our experimental approach reflects this reality, and unlike standard approaches, we choose (which makes sense for the approximation considered) to initiate the flow in the IR by inducing the theory in the UV.
Although this approach is paradoxical from the point of view of the very meaning of the RG (a semigroup), it is not totally exotic either.
It is, for example, the point of view adopted in the literature devoted to the problem of ``asymptotic safety'' in quantum gravity~\cite{Eichhorn:2018yfc}, aiming to prove that quantum gravity is well-defined in the UV, even if it is not (perturbatively) just-renormalizable. 

\begin{figure}[t]
    \centering
    \begin{subfigure}[t]{0.49\linewidth}
        \centering
        \includegraphics[width=\linewidth]{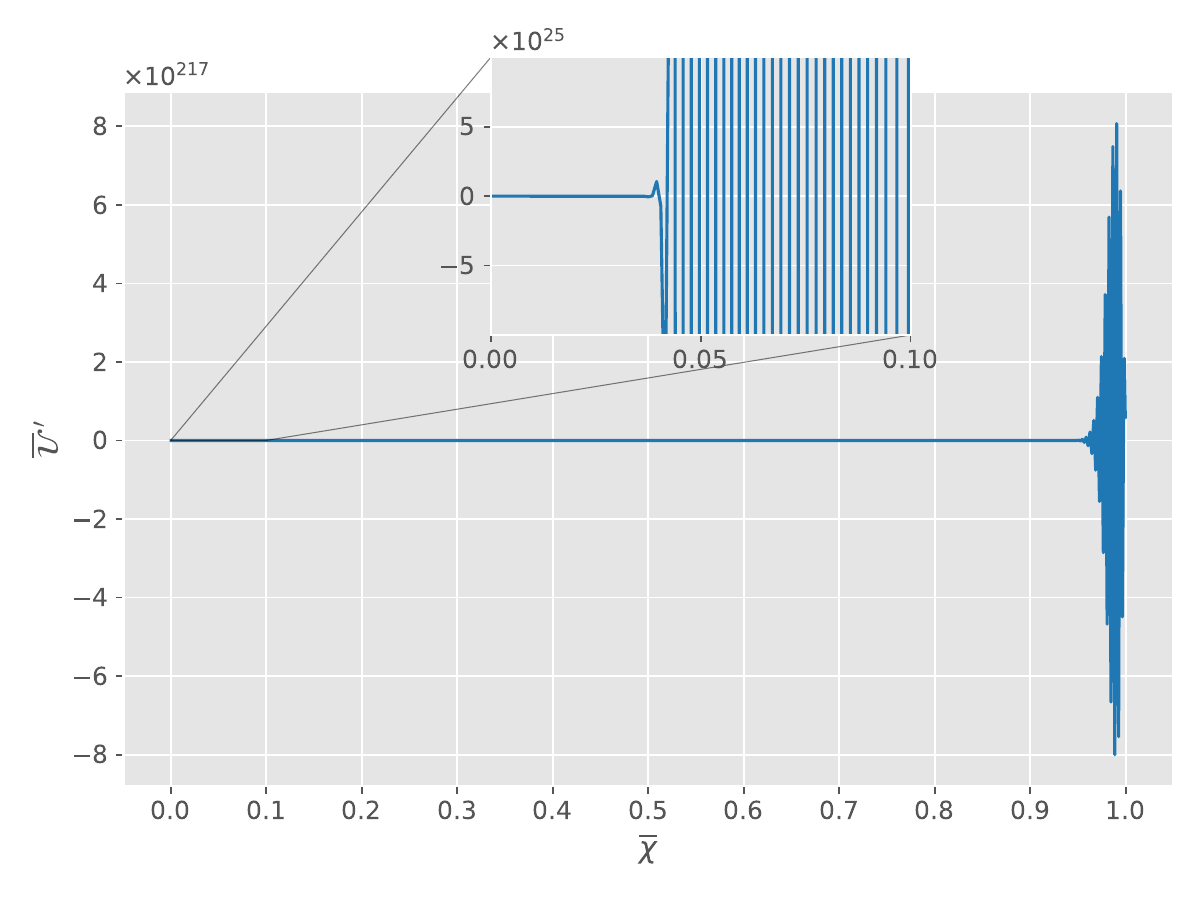}
        \caption{Variation at $k = 0.02$.}
    \end{subfigure}
    \hfill
    \begin{subfigure}[t]{0.49\linewidth}
        \centering
        \includegraphics[width=\linewidth]{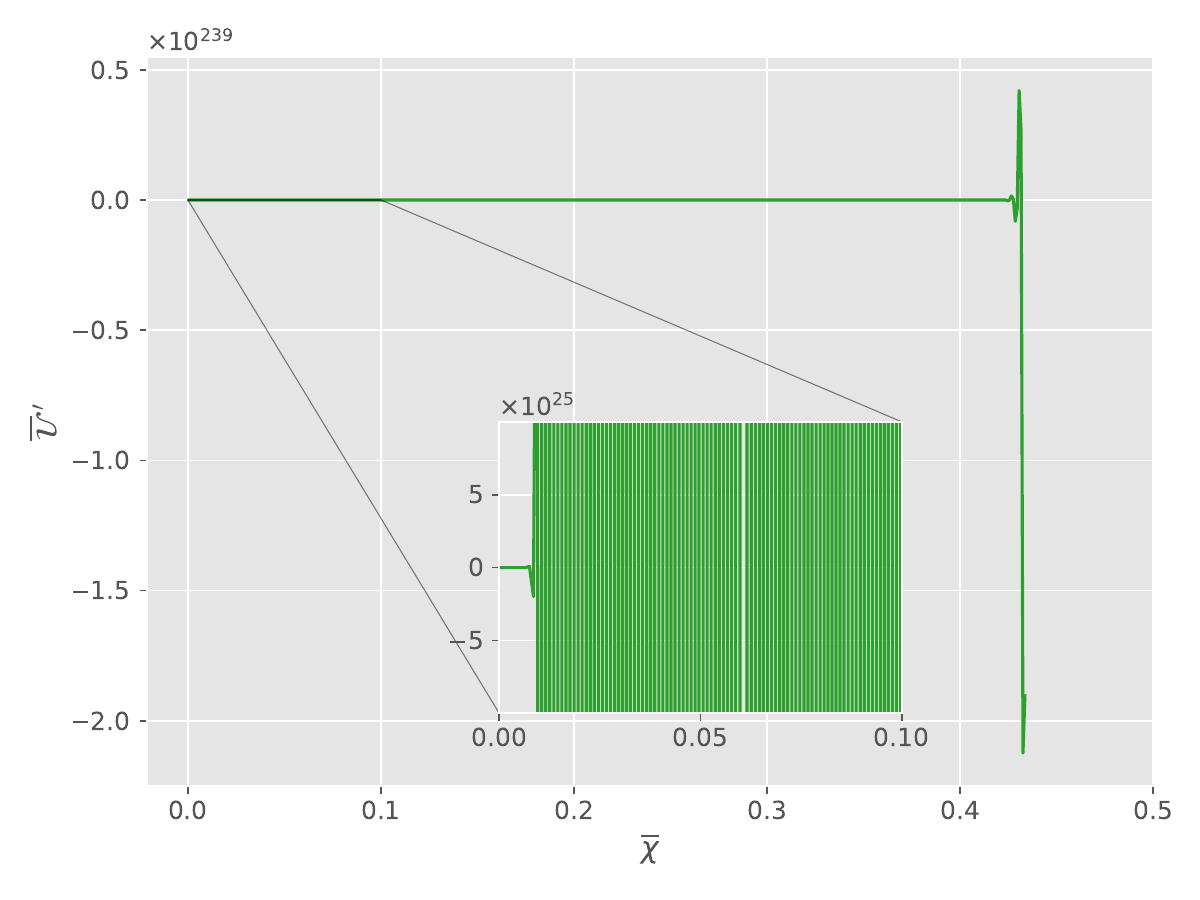}
        \caption{Variation at $k = 0.06$.}
    \end{subfigure}
    \\
    \begin{subfigure}[t]{0.49\linewidth}
        \centering
        \includegraphics[width=\linewidth]{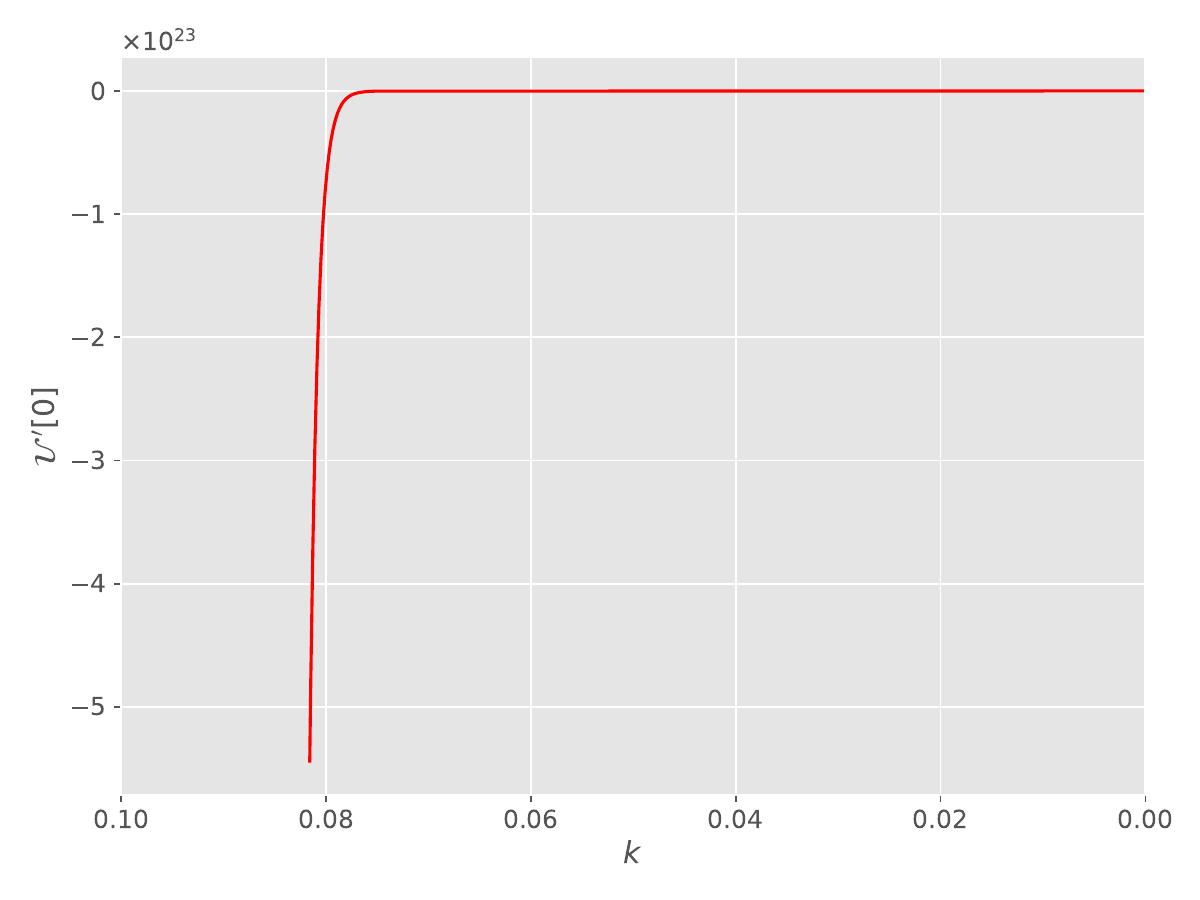}
        \caption{Back-evolution of $\overline{\mathcal{U}}^{\,\prime}[0]$.}
    \end{subfigure}
    \hfill
    \begin{subfigure}[t]{0.49\linewidth}
        \centering
        \includegraphics[width=\linewidth]{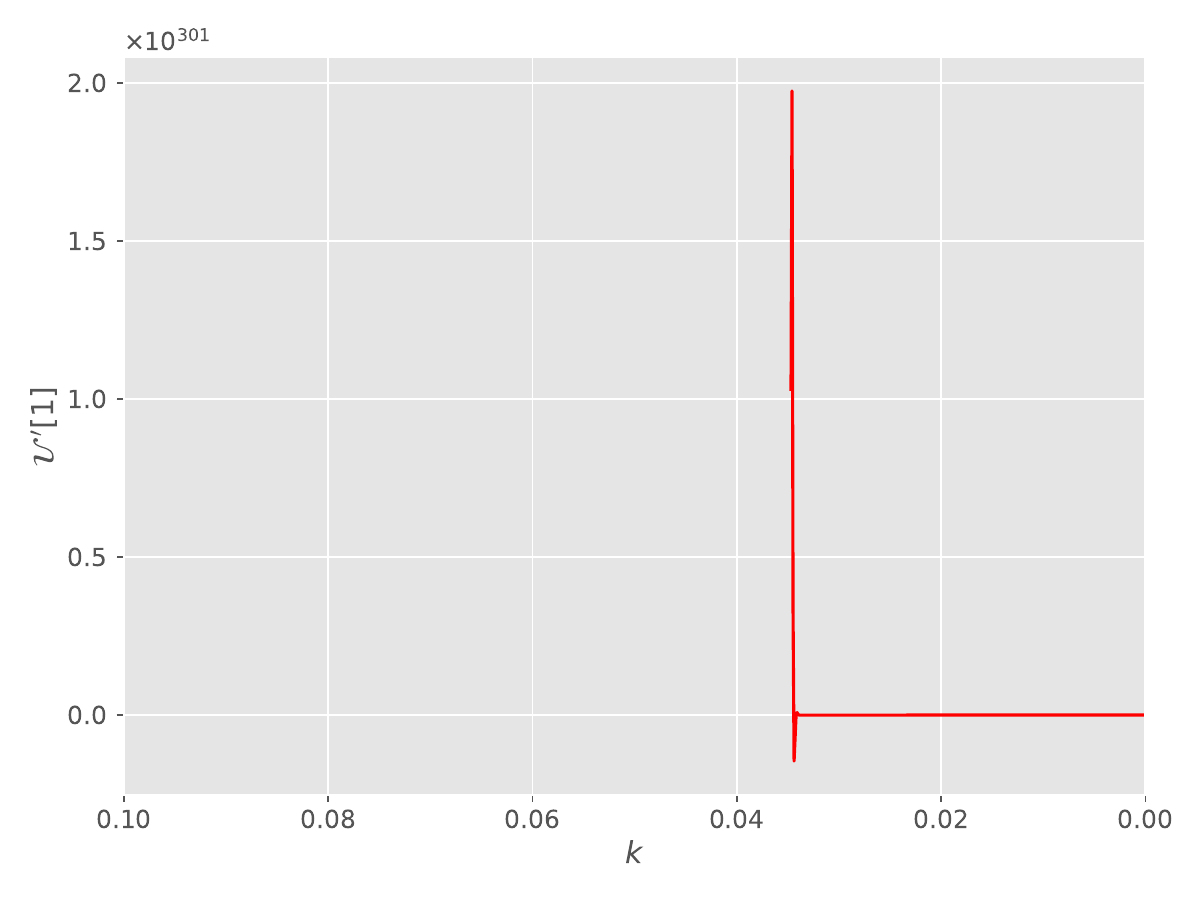}
        \caption{Back-evolution of $\overline{\mathcal{U}}^{\,\prime}[1]$.}
    \end{subfigure}
    \caption{%
        Typical variation of the potential and its back-evolution for the analytical MP law.
        The initial conditions are the following:
        $\bar{\mu}_1 = \bar{\mu}_3 = \bar{\mu}_4 = 0.0$,
        $\bar{\mu}_2 = 1.0$. Arbitrary large and rapid oscillations are observed for the back-evolved potential, after some RG steps.
    }
    \label{back-evolution-MP}
\end{figure}

\subsection{Results and comments}

First, we consider the case of the MP spectrum.
The corresponding $\rho$'s shape is pictured in Figure~\ref{inverse_MP}, and Figure~\ref{back-evolution-MP} shows the typical behaviour of the back-evolution of the effective potential, with general initial condition \eqref{eq:pot_initial}.

\begin{figure}[t]
    \centering
    \begin{subfigure}[t]{0.49\linewidth}
        \centering
        \includegraphics[width=\linewidth]{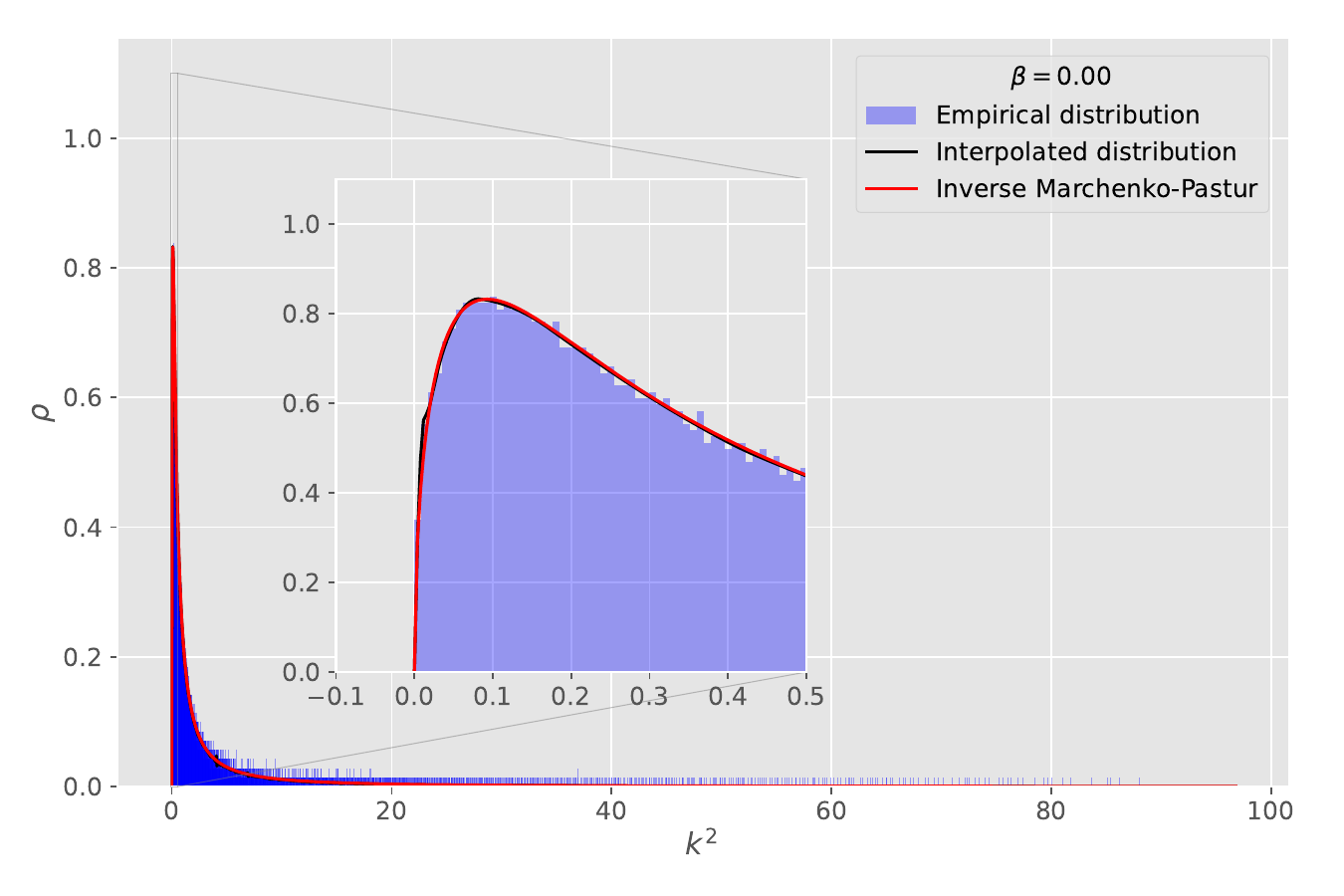}
        \caption{%
            Noisy ($\beta = 0$) empirical distribution.
        }
    \end{subfigure}
    \hfill
    \begin{subfigure}[t]{0.49\linewidth}
        \centering
        \includegraphics[width=\linewidth]{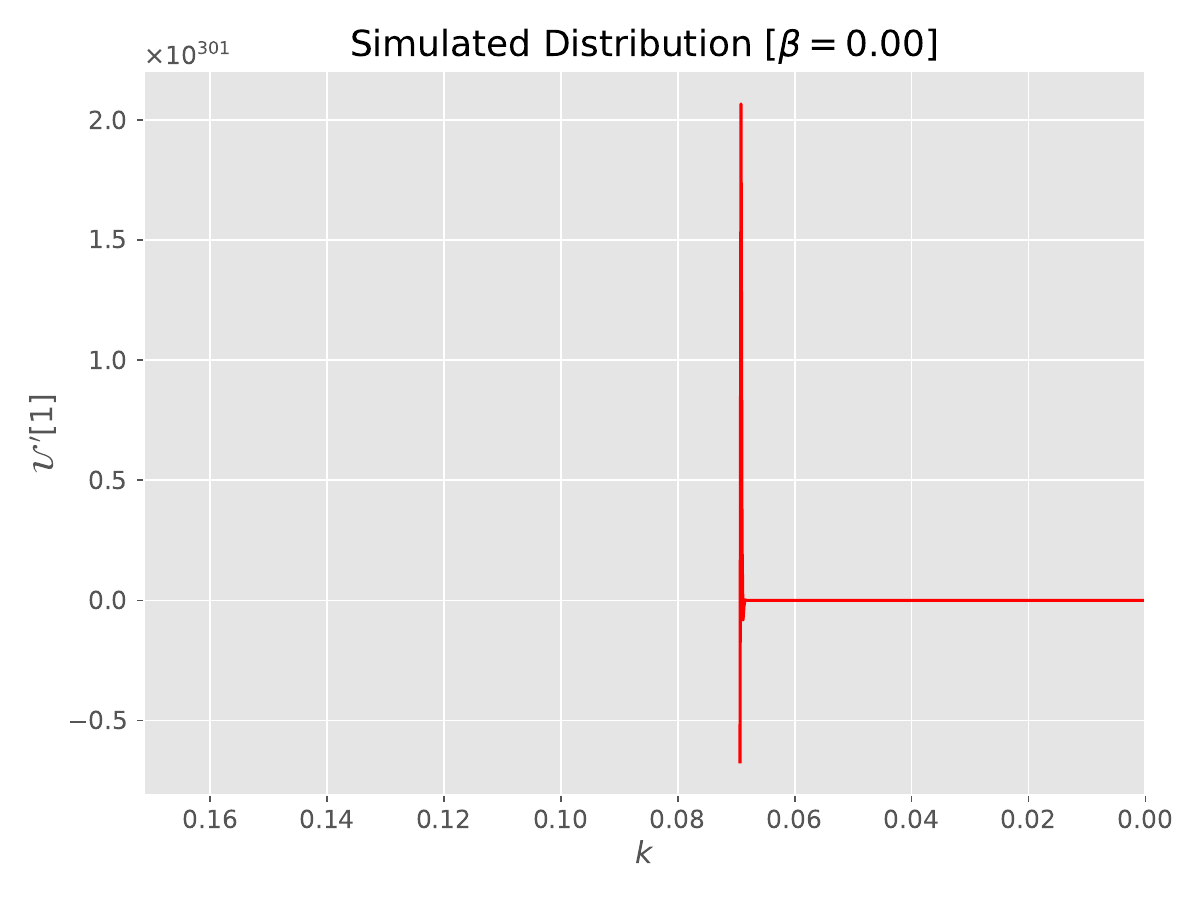}
        \caption{%
            RG flow for $\beta = 0$.
        }
    \end{subfigure}
    \caption{%
        The behaviour of the RG flow in the case of the empirical distribution for $\beta = 0$ is extremely similar to the one obtained from the analytical MP law.
        The absence of values for $k \gtrsim 0.07$ signals the divergence of the values in the Python simulation.
    }
    \label{fig:empirical_noisy}
\end{figure}

Recall that, asymptotically in the IR, the power counting for the MP law reduces to the one of a $3$-dimensional field theory, and that all interactions higher than sextic are irrelevant (see Figure~\ref{canonical-dimensions}).
The behaviour we observe is reminiscent of that found in classical models of disordered systems, such as $p$-spin models, whose functional renormalization group has recently been considered~\cite{Lahoche:2021tyc,lahoche2022functional} in a temporal approach equivalent to that considered in this article.
The authors observed the same type of divergence at finite $k$, which can be physically interpreted as the failure of the assumption that the system is in equilibrium dynamics.
Indeed, this assumption is implicitly made in the construction of the MSR path integral \eqref{MSRZ} because we took the origin of time for $t = -\infty$, thanks to the expected time reversal symmetry.
In the two previous references, the authors especially confronted these assumptions, and showed that finite time singularities are truly related with a breakdown of the time reversal symmetry (focusing on the breakdown of the underlying supersymmetry in~\cite{Lahoche:2021tyc} and through a $2$PI formulation in~\cite{lahoche2022functional}).
Hence, we recover in this model what we could expect from previous investigations: the system fails to reach equilibrium for general initial conditions in the deep IR.\footnote{%
    We could say for \emph{almost all} initial conditions.
    It is indeed possible that the situation is different for some of them, as we saw in~\cite{lahoche2022functional} (and reference therein).
    However, we have not yet been able to identify any such conditions.
}

It is also worth noting, for the following discussion, that these conclusions are not specific to the $P,\, N \to \infty$ limit.
They remain true even when these parameters are finite, provided they are sufficiently large.
Figure~\ref{fig:empirical_noisy} illustrates this point, and the parameter $\beta$ set to zero means the no signal limit (see the convention given in the introduction, though this point will be clarified in a few moment).
Once again, we recover the same phenomena: a finite scale divergence reminiscent of a weak ergodicity breaking.

\begin{figure}[t]
    \centering
    \begin{subfigure}[t]{0.49\linewidth}
        \centering
        \includegraphics[width=\linewidth]{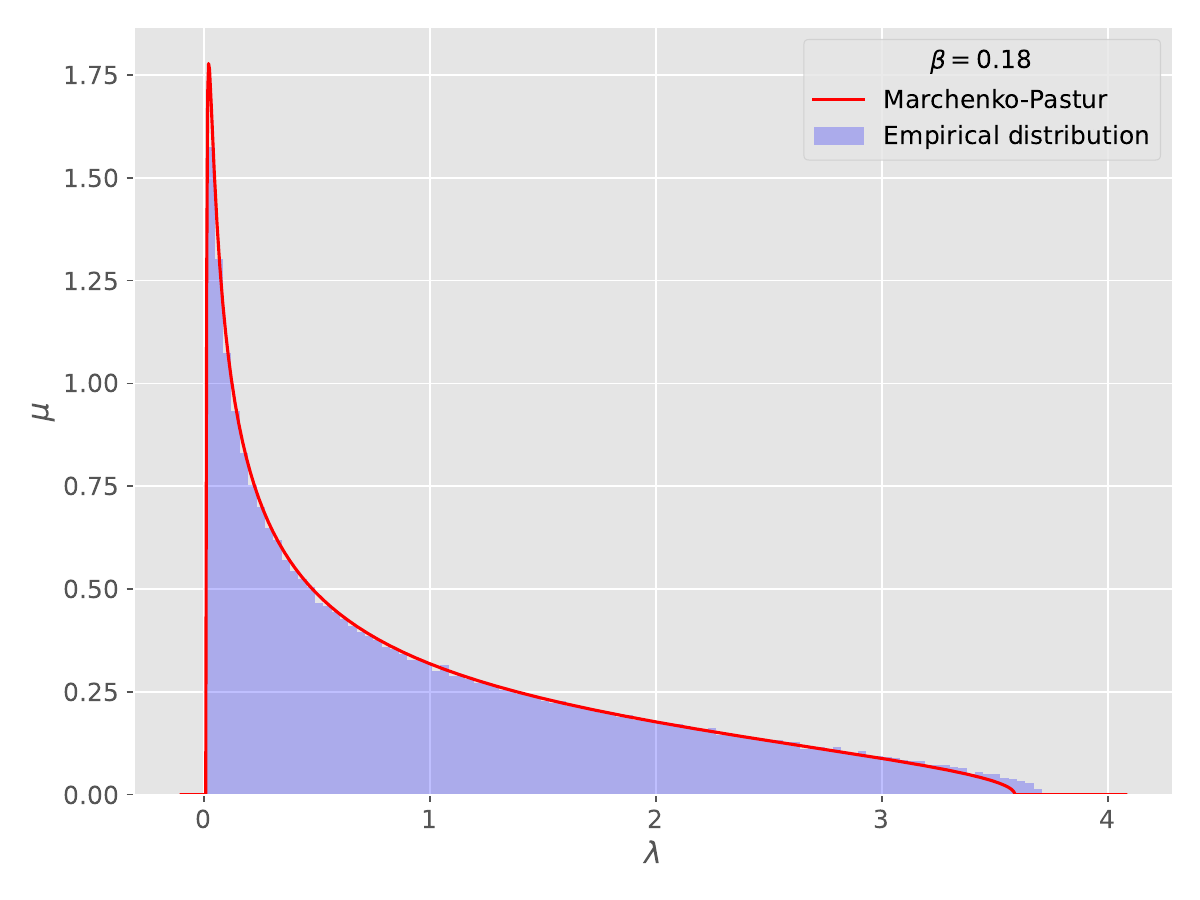}
        \\
        \includegraphics[width=\linewidth]{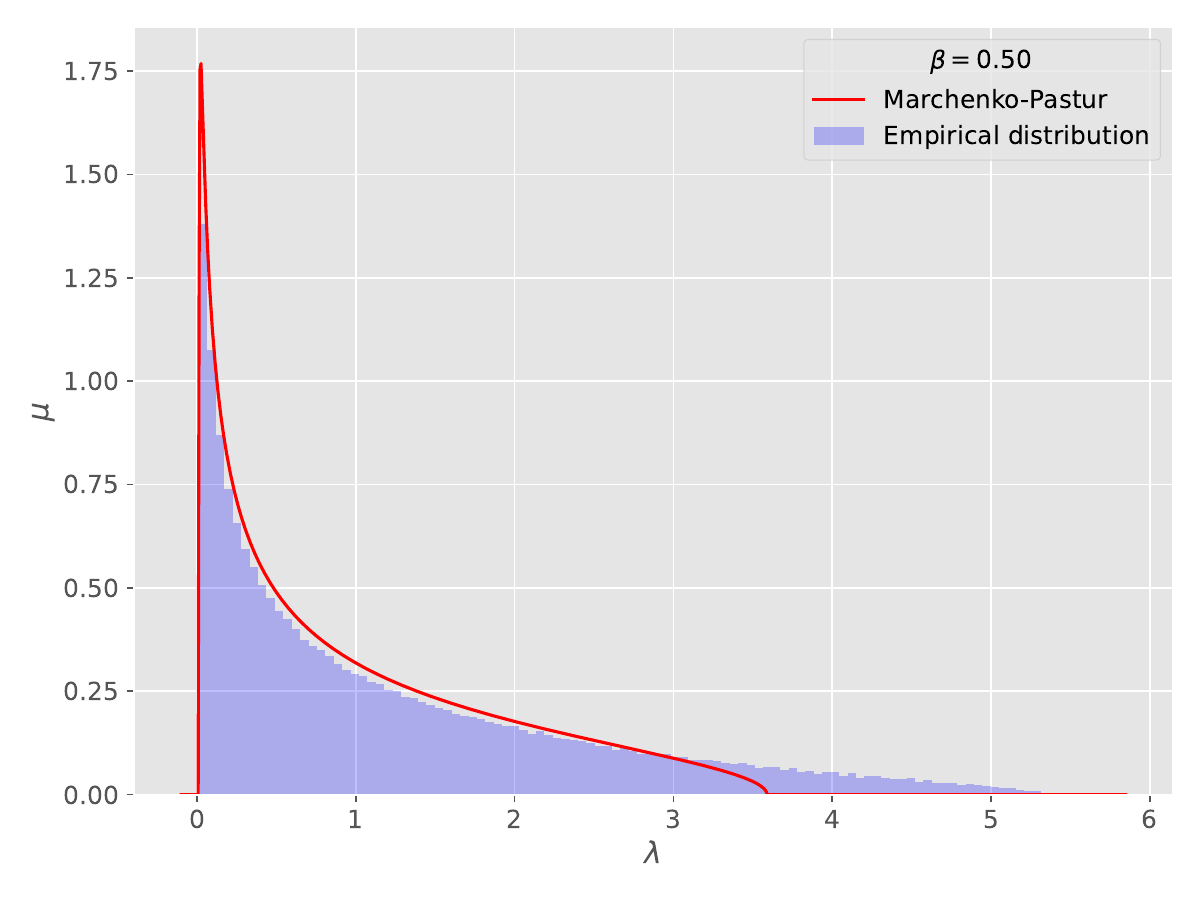}
        \caption{%
            Simulated empirical distribution.
        }
    \end{subfigure}
    \hfill
    \begin{subfigure}[t]{0.49\linewidth}
        \centering
        \includegraphics[width=\linewidth]{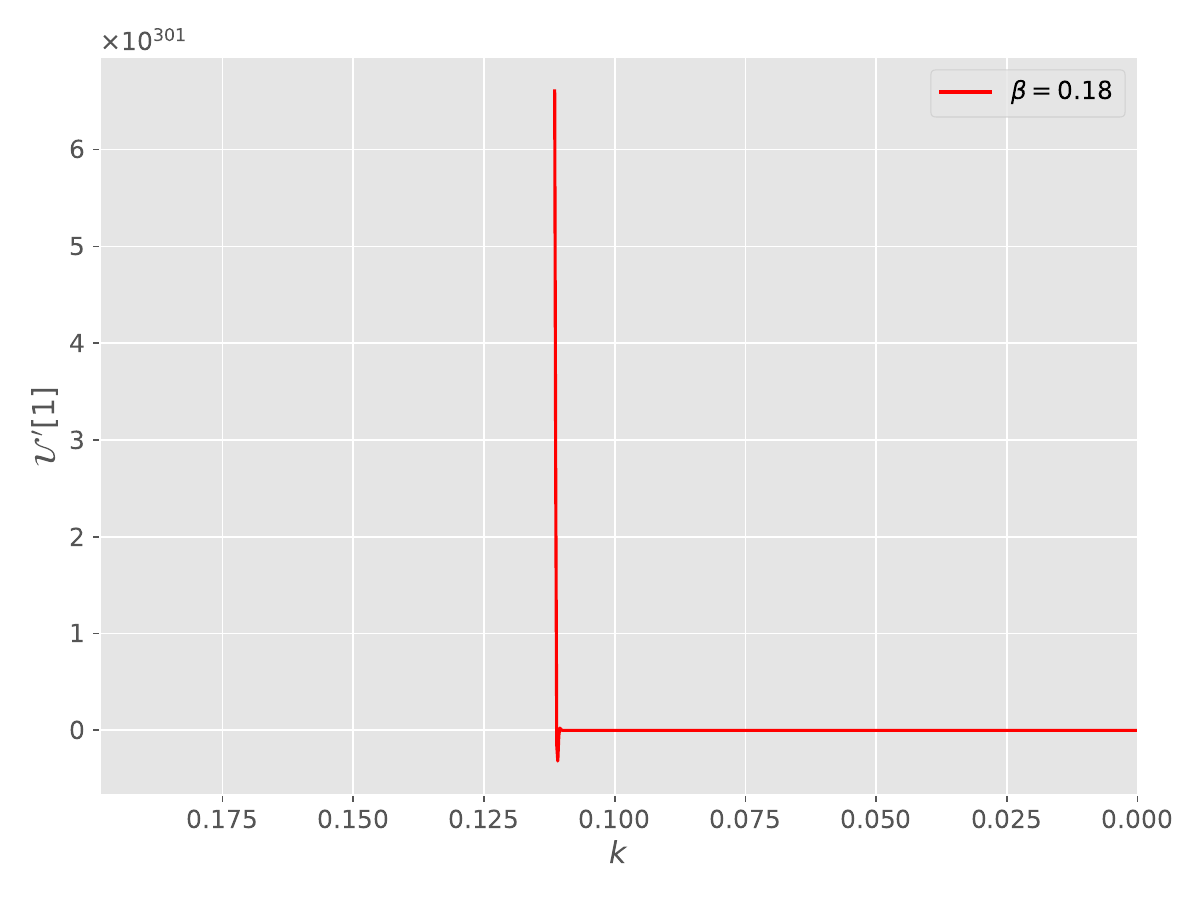}
        \\
        \includegraphics[width=\linewidth]{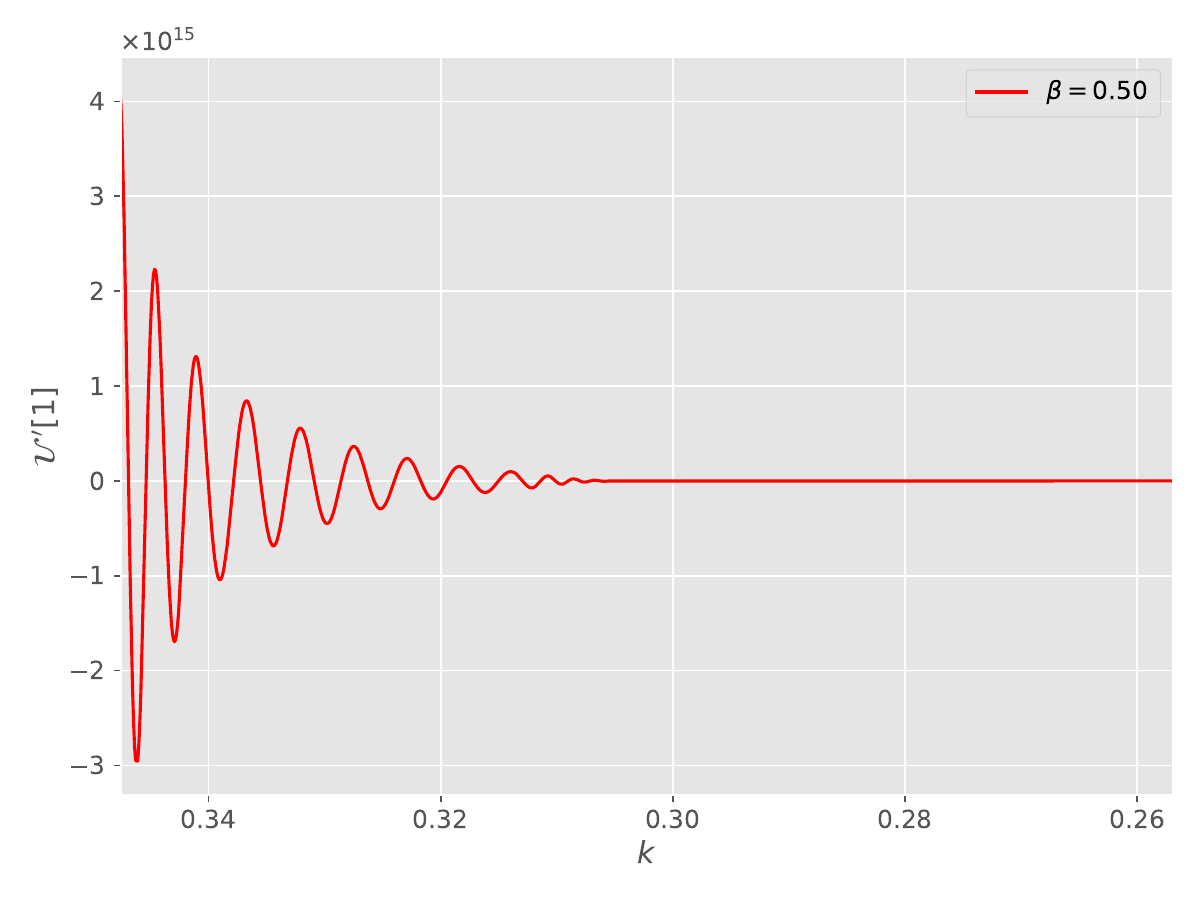}
        \caption{%
            RG flow.
        }
    \end{subfigure}
    \caption{%
        Empirical spectrum with increased level of signal (on the left), and the corresponding back evolution of derivative of the potential at the (arbitrary) value $\bar{\chi}=1$.
    }
    \label{fig:simulated_flow}
\end{figure}

Figure~\ref{fig:simulated_flow} shows some results, for different values of the parameter $\beta$. 
On the left, we can see the empirical distribution for a given draw of $Z$, and on the right we can show the corresponding RG flow (with the same initialization as on the Figure~\ref{back-evolution-MP}).
As long as $\beta$ remains small, the results remain qualitatively those obtained previously for a purely noisy signal; there is, however, a slight delay in the explosion of the RG flow.
For $\beta$ large enough, the behaviour of the RG flow is very different.
Although the potential takes large absolute values, it does not seem to diverge, and its derivative also remains regular.
Thus, our initial hypothesis seems to be confirmed: \emph{a sufficiently large signal favours a return to equilibrium}, or at least does not reject this hypothesis, which is central to the construction of the MSR partition function.
This qualitative change in the behaviour of the flow can be used to define a signal detectability threshold.

\subsection{Exploring the phase space}

\begin{figure}[t]
    \centering
    \begin{subfigure}[t]{0.49\linewidth}
        \centering
        \includegraphics[width=\linewidth]{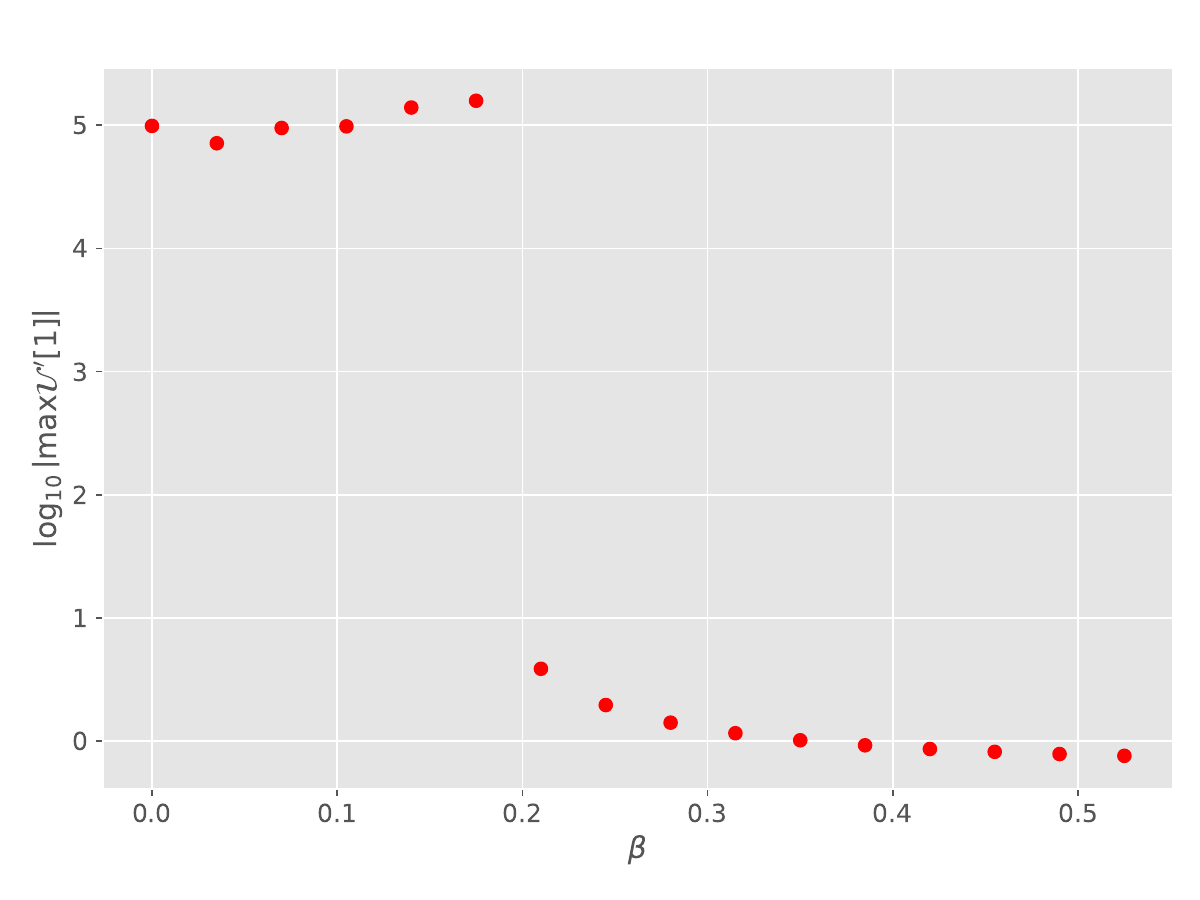}
        \caption{%
            Limit $T \to 0$.
        }
    \end{subfigure}
    \hfill
    \begin{subfigure}[t]{0.49\linewidth}
        \centering
        \includegraphics[width=\linewidth]{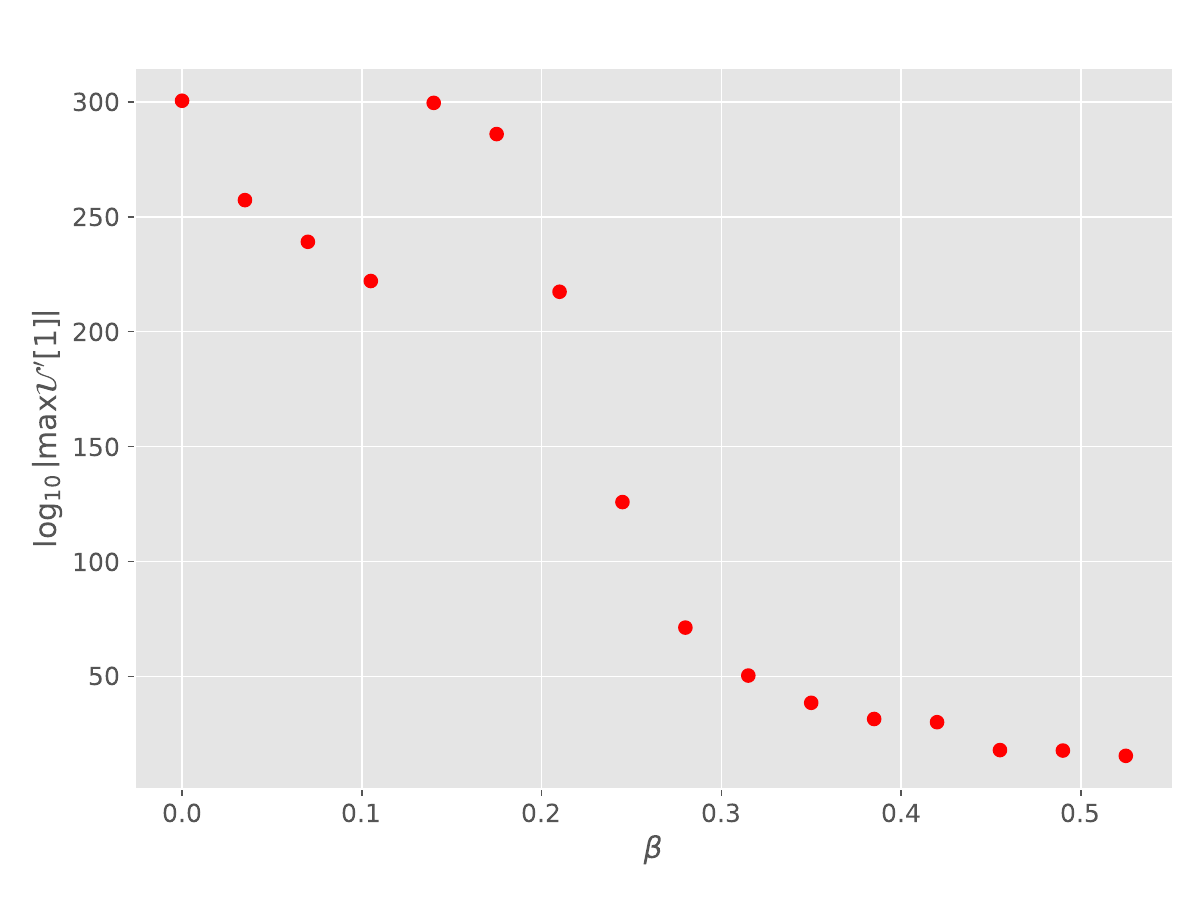}
        \caption{%
            $T = 0.50$.
        }
    \end{subfigure}
    \\
    \begin{subfigure}[t]{0.49\linewidth}
        \centering
        \includegraphics[width=\linewidth]{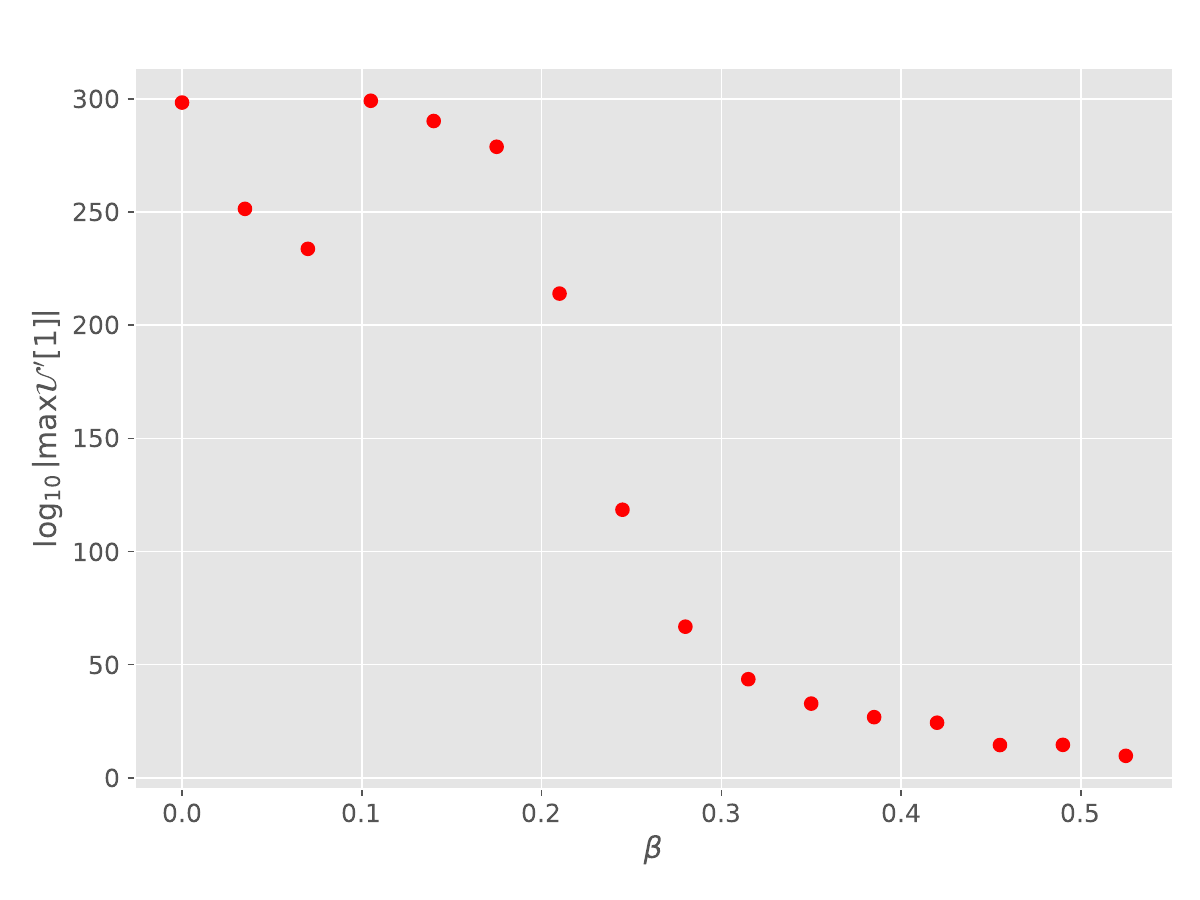}
        \caption{%
            $T = 1.10$.
        }
    \end{subfigure}
    \hfill
    \begin{subfigure}[t]{0.49\linewidth}
        \centering
        \includegraphics[width=\linewidth]{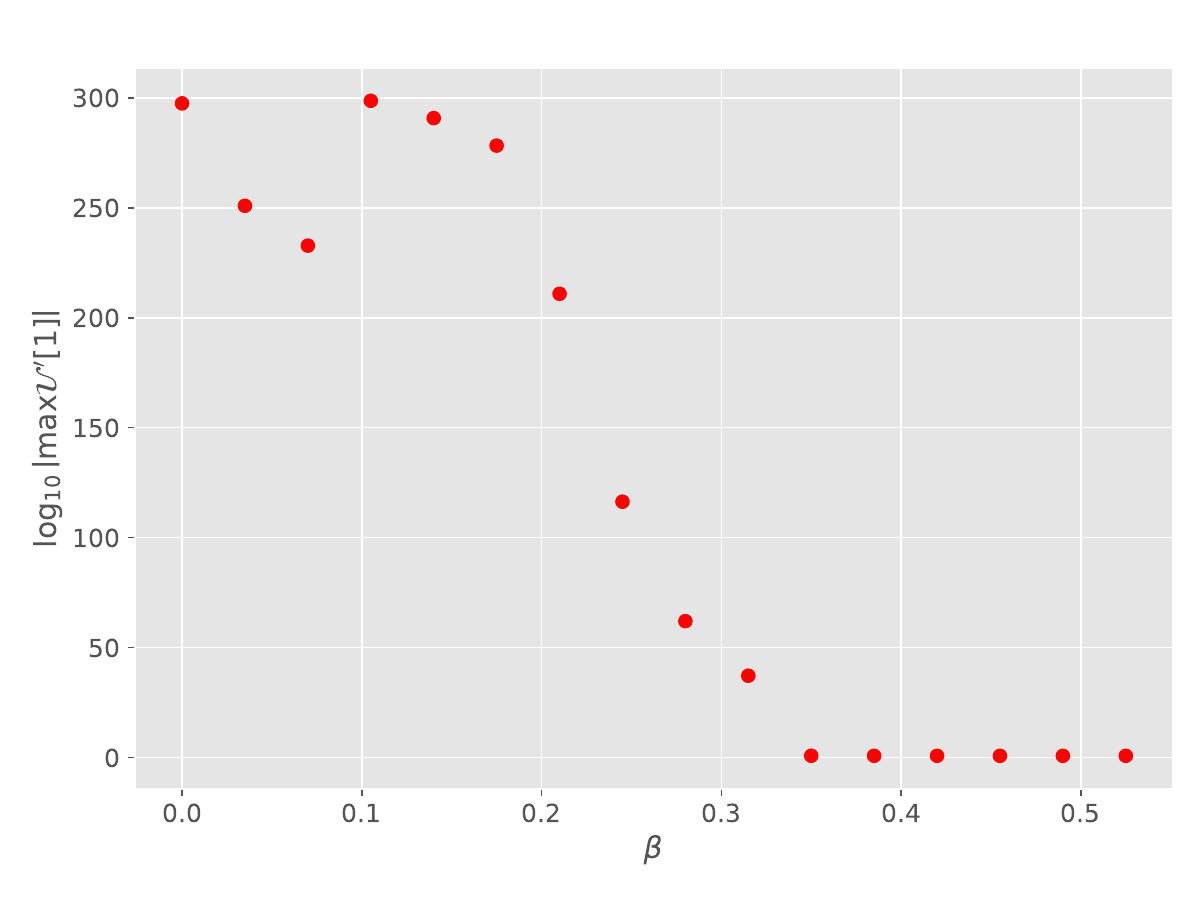}
        \caption{%
            $T = 1.55$.
        }
    \end{subfigure}
    \caption{%
        We show the behaviour of the (suitably normalized) absolute value for the maximum of the derivative of the potential at the point $1$ with respect to $\beta$ and for different temperature.
    }
    \label{fig:temperature_dep}
\end{figure}

\begin{figure}[t]
    \centering
    \begin{subfigure}[t]{0.49\linewidth}
        \centering
        \includegraphics[width=\linewidth]{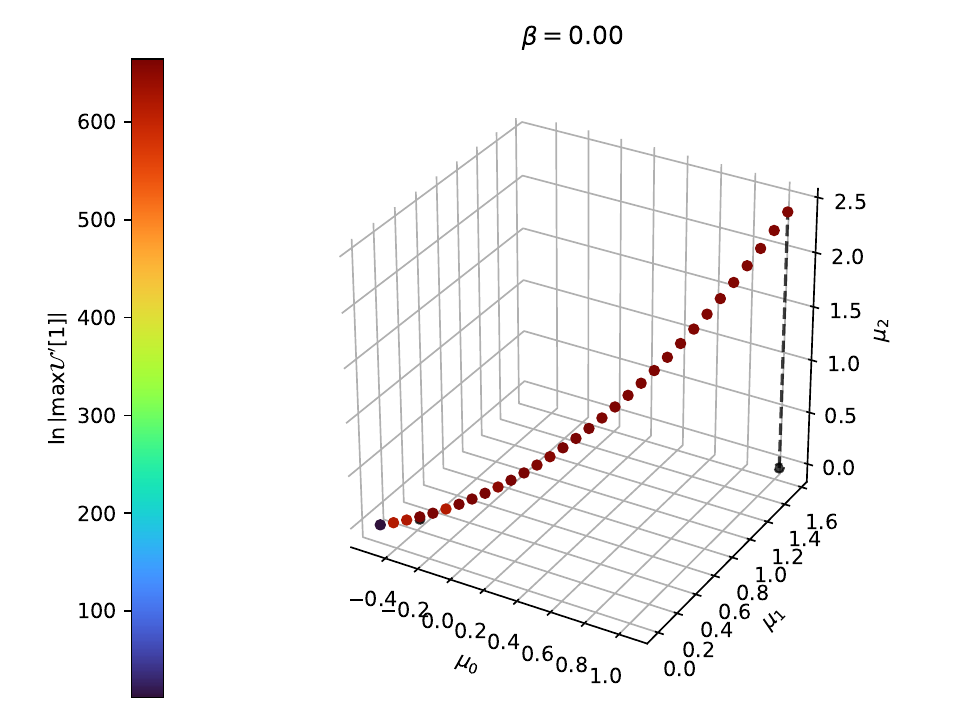}
        \caption{%
            Phase diagram at $\beta = 0.00$.
        }
    \end{subfigure}
    \hfill
    \begin{subfigure}[t]{0.49\linewidth}
        \centering
        \includegraphics[width=\linewidth]{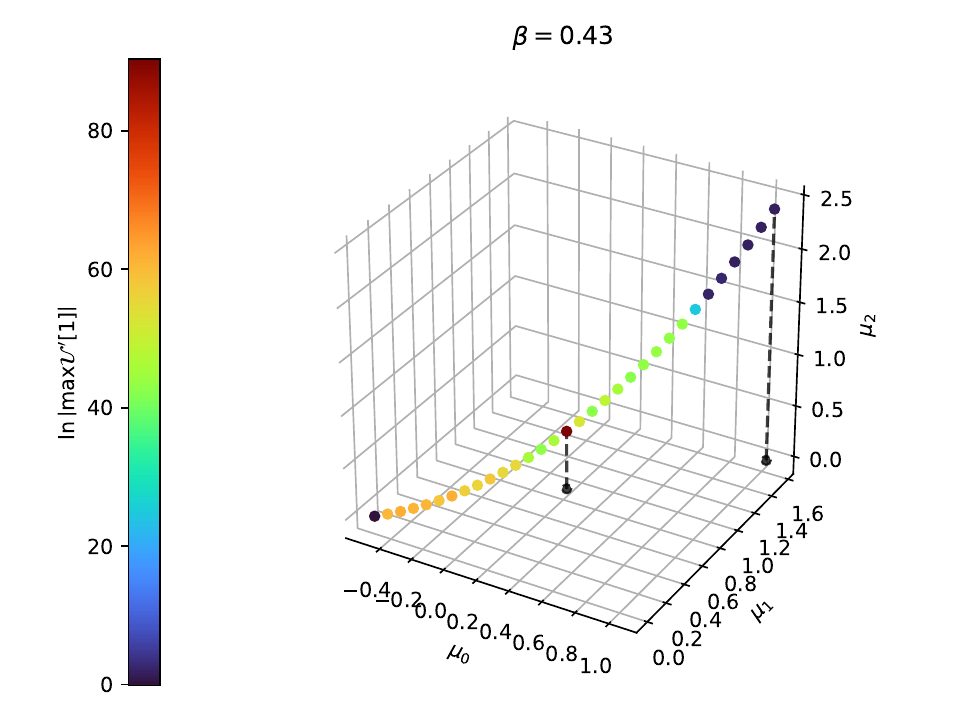}
        \caption{%
            Phase diagram at $\beta = 0.43$.
        }
    \end{subfigure}
    \\[1em]
    \begin{subfigure}[t]{0.6\linewidth}
        \includegraphics[width=\linewidth]{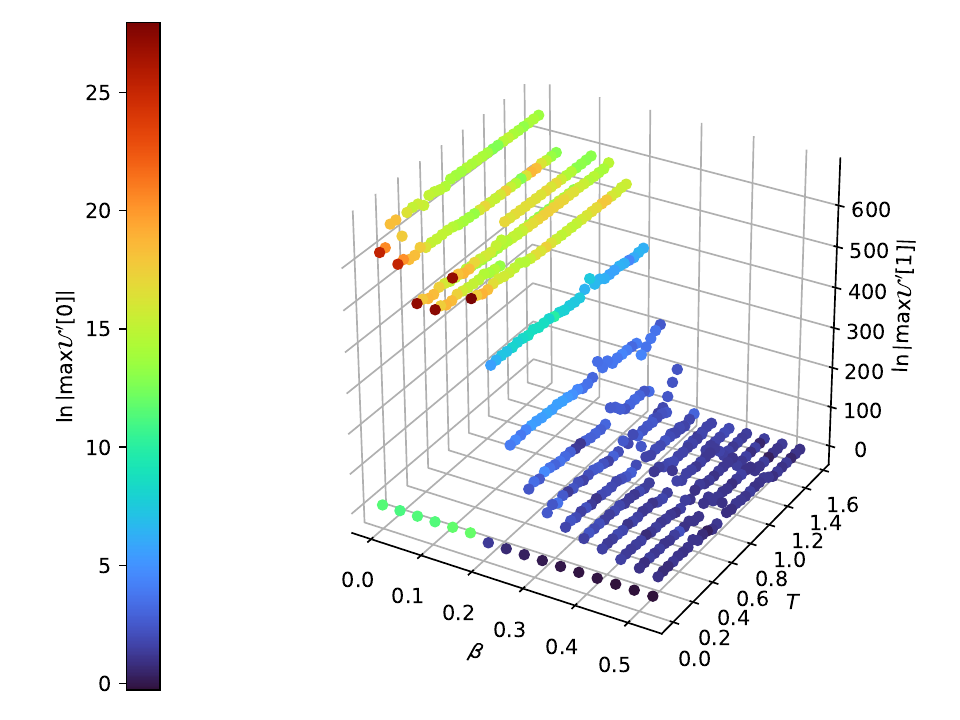}
        \caption{%
            Maxima of the potential for different values of $\beta$, in different points of the phase space.
        }
    \end{subfigure}
    \caption{%
        We show the maximum values of the derivative of the potential at the value $\bar{\chi} = 1$ in the parameter space.
    }
    \label{fig:phase_space_exploration}
\end{figure}

In what follows, we choose a more convenient representation of the initial conditions in the IR, based on a single parameter that we will identify (after redefining the fields) as the (renormalized) temperature $T$:
\begin{equation}
    \bar{\mu}_1 = \frac{T-T_0}{T},
    \qquad
    \bar{\mu}_{n>1} = T^{n-1}.
    \label{parametrizationIRT}
\end{equation}
In this parametrization, the temperature $T_0$ is arbitrary, and we set $T_0 = 0.5$ as a ``test'' value in the experiments (in this scenario, averaged over 10 different realizations each).
In Figure~\ref{fig:temperature_dep}, we plot the logarithm of the absolute value of the maximum of the derivative of the potential as a function of the value of $\beta$, for different temperatures.
We can clearly identify a transition between two regimes, around the value $\beta_c \simeq 0.2$.
Notice that this value does not seem to depend significantly on the $T_0$ value as well as on the value of $\bar{\chi}$ selected for the simulation : we were able to verify the statement numerically.
It is worth noting that, although we had identified the existence of a critical $\beta_c$ value for signal detection in our previous work~\cite{LahocheSignal2022}, we had not yet been able to estimate it.
Finally, let us point out that while the figures appear similar for different temperatures, the $T \to 0$ limit is singular, and the transition to the critical value $\beta_c$ seems more abrupt.
This limit would physically correspond to the case of a totally uncorrelated noise; this phenomenon deserves to be completed by a finer physical analysis of the corresponding model, and we mention its results as an opening towards further investigations.
In the same way, the above analysis does not yet provide a precise answer to the other question in the introduction, and to which we have already provided some elements of an answer in our previous articles.
Namely, the physically motivated position of a boundary between signal and noise.
We expect the delocalization of eigenvectors to induce a mixture between the ``information'' and ``noise'' parts of the signal; so we could, as we explained in the introduction, hope to determine a boundary on the spectrum at which we expect to find more information than noise.
Here, the search for this boundary using the formalism developed in this article is a work in progress and will be the subject of future works. 

\begin{figure}[t]
    \centering
    \includegraphics[width=0.7\linewidth]{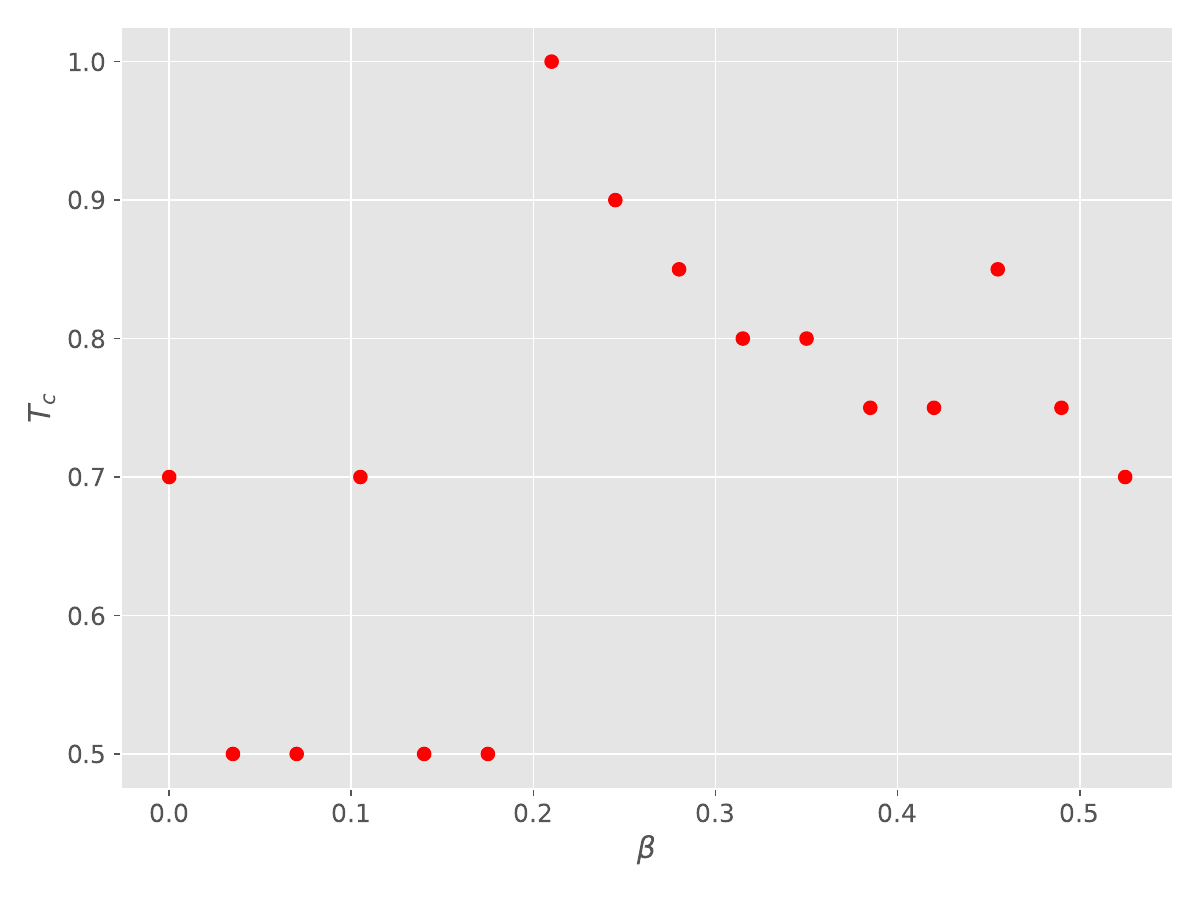}
    \caption{%
        The behaviour of $T_c$ as a function of the signal strength $\beta$.
    }
    \label{fig:critical_temp}
\end{figure}

We conclude the section by mentioning other results that seem to confirm the previous ones.
Starting with some initial conditions, we show on the Figure~\ref{fig:phase_space_exploration} the maximal values of the derivative on the potential for two values of $\beta = 0$ and $\beta = 0.43$.
According to the previous results, the values taken by the potential with vanishing $\beta$ are very large, and then diverge at almost every point of the phase space.
The situation is different for $\beta = 0.43$, where the potential takes large, though finite values.\footnote{%
    According to our machine limit, which does not distinguish numbers larger than $10^{308}$ with infinity.
}
This is indeed true for almost every point on the set represented in the figure, except for a point where the potential seems to take a large value.
The same phenomenon occurs for different values of $\beta$, as the last plot in Figure~\ref{fig:phase_space_exploration} shows, using parametrization \eqref{parametrizationIRT} for the initial conditions.
We can observe that, for each value of $\beta$, there exists a value for which the potential forms a ``cusp'', for a given $T_c$ value of the temperature.
The value of this ``critical'' temperature depends on $\beta$, and the Figure~\ref{fig:critical_temp} shows explicitly its dependency. The value of the critical temperature shows the same kind of qualitative change as we had previously observed, in the vicinity of the critical value $\beta_c\simeq 0.2$. Here again, we see that the net effect of the signal is to delay the transition point, by increasing the effective value of the critical temperature, for a given parameterization fixed in the IR. 

\section{Conclusion}\label{sec4}

This article continues recent work by the same authors, summarized in~\cite{LahocheSignal2022}, aimed at exploiting the functional renormalization group for signal detection when the latter is hidden within highly noisy degrees of freedom (low signal-to-noise ratio).
This problem has yet to be solved, or solved efficiently, by current methods.
Our aim is to identify not only a detection threshold, but also a characteristic spectrum scale for distinguishing a ``noisy'' sector.
In this article, we have considered stochastic field theory and investigated the relationship between the presence of a signal and the return to equilibrium.
The net result is the existence of a transition between two clearly identified regimes, a first regime where the system never reaches equilibrium (noisy regime), and a regime where the equilibrium condition can be maintained, when the signal strength is large enough.
We were thus able to give an estimate of the critical $\beta_c$ value at the detection threshold, which we were unable to do in previous investigations.
However, we still have a long way to go in understanding the physics of these flows.
For example, we have not studied the characteristics of the potential but only its divergences, nor have we finely analysed the relevance of our approximations (LPA) when we deviate from the IR.
Finally, we have not gone any further in estimating the boundary with noisy degrees of freedom, but this question is the subject of ongoing work, and will be dealt with in the direct aftermath of this work, which should be considered as preliminary.

\section*{Acknowledgments}

This project has received funding from the European Union's Horizon 2020 research and innovation program under the Marie Skłodowska-Curie grant agreement No 891169.
This work is supported by the National Science Foundation under Cooperative Agreement PHY-2019786 (The NSF AI Institute for Artificial Intelligence and Fundamental Interactions, \url{http://iaifi.org/}).
The simulations presented in this article were performed on the FactoryIA supercomputer, financially supported by the Île-De-France Regional Council, and the Engaging cluster at the MGHPCC facility.
We also acknowledge the European COST Action \href{https://www.cost.eu/actions/CA22130/}{CA22130} for the profitable exchanges on the use of the techniques in this article for data analysis.

\pagebreak
\printbibliography[heading=bibintoc]
\end{document}